\documentclass[12pt]{article}

\usepackage{epsf}

\newcommand{\vev}[2]{\langle {#2}|\, {#1} \,| 0\rangle}

\begin{document}
%-------------------------------------------------------------------

\begin{flushright}

SLAC-PUB-8822 \\
DESY-01-060 \\
hep-ph/0105194 \\
May 2001
\end{flushright}

\renewcommand{\thefootnote}{\fnsymbol{footnote}}

\begin{center}
\vskip 4.5\baselineskip
\textbf{\Large New ways to explore factorization in $b$
decays\,\footnote{Work supported by Department of Energy contract
DE--AC03--76SF00515.}}
\vskip 3.5\baselineskip
M.~Diehl$^{1}$ and G.~Hiller$^{2}$
\vskip \baselineskip
1. Deutsches Elektronen-Synchroton DESY, 22603 Hamburg, Germany \\
2. Stanford Linear Accelerator Center, Stanford University,
   Stanford, CA 94309, U.S.A.
\vskip 8.5\baselineskip
\textbf{Abstract}\\[0.5\baselineskip]
\parbox{0.9\textwidth}{We propose to study factorization breaking
effects in exclusive $b$ decays where they are strongly enhanced over
the factorizing contributions. This can be done by selecting
final-state mesons with a small decay constant or with spin greater
than one. We find a variety of decay modes which could help understand
the dynamical origin of factorization and the mechanisms responsible
for its breaking.}
\vskip 2.5\baselineskip
\vskip 1.5\baselineskip
\end{center}

\renewcommand{\thefootnote}{\arabic{footnote}}
\setcounter{footnote}{0}

\newpage

% end of title page

%%%%%%%%%%%%%%%%%%%%%%%%%%%%%%%%%%%%%%%%%%%%%%%%%%%%%%%%%%%%%%%%%%%
\section{Introduction}
\label{sec:intro}
%%%%%%%%%%%%%%%%%%%%%%%%%%%%%%%%%%%%%%%%%%%%%%%%%%%%%%%%%%%%%%%%%%%

An outstanding task in heavy-flavor physics is to understand the
strong interaction effects in exclusive weak decays of hadrons
containing a $b$-quark. For many decay channels, such an understanding
is a precondition for gaining information on the quark mixing matrix
or on physics beyond the standard model. In addition, the dynamics of
quarks, gluons, and hadrons in the presence of a large mass $m_b$ is
interesting in QCD by its own right.

Introduced in \cite{Bauer:1987bm}, the concept of factorization has
been one of the most successful tools in this respect, providing fair
agreement between theory and data for many channels. In other cases,
factorization in its most naive version fails when compared with
experiment, and there have been several phenomenologically motivated
improvements over its original form \cite{Neubert:1997uc,Ali:1998eb}.

There are several dynamical arguments why and where factorization
should be valid. One is based on the large $N_c$ limit of QCD
\cite{Buras:1986xv}, whereas a different line of approach builds on
the color transparency phenomenon \cite{Bjorken:1989kk}. More
recently, the framework of QCD factorization has implemented the color
transparency argument in the language of perturbation theory and power
counting in $1/m_b$ \cite{Beneke:1999br,Beneke:2000ry,Beneke:2001ev}.

In these approaches it is understood that there are corrections to
naive factorization, which are suppressed in a small parameter such as
$1/N_c$, or $\alpha_s$ and $\Lambda_{\mathrm{QCD}} /m_b$. A more
quantitative understanding of their size is crucial in order to assess
for which channels and to which precision the factorization concept
can be applied. There are also scenarios where factorization in the
sense of \cite{Bauer:1987bm} does not appear as a limit when a small
parameter vanishes, and where conceptually factorization breaking
terms in the decay amplitude may be as large as the factorizing ones.
An example is the perturbative hard scattering (PQCD) approach
\cite{Keum:2001wi}. In view of such controversies, and given that
present day theory can at best estimate the size of most
nonfactorizing contributions, quantitative tests of factorization in
the data are of great importance.

We propose here to study decay channels where the factorizing
contributions to the amplitude are small or zero for symmetry
reasons. In such a situation nonfactorizing contributions, which would
otherwise be suppressed, have a chance to be clearly visible. The
measurement of the corresponding decay rates can thus give rather
direct information on their size, and the comparison of different
channels may give indications on the relevant dynamical mechanisms.

Our suggestion is to choose decay channels whose flavor structure is
such that a selected meson $X$ must be emitted from the weak current
mediating the $b$-quark decay. Taking then a meson which has as very
small decay constant, the factorizable contributions to the decay are
suppressed. A second possibility is to consider mesons $X$ with spin
$J\ge 2$. A tensor meson for instance cannot be produced from a
decaying $W$ boson, which has spin 1, unless there are interactions
involving the other hadrons in the decay process. We will find a
variety of decay channels where these ideas can be realized, which
will allow us to address different issues related to factorization and
its breaking.

The organization of this paper is as follows. In Sect.~\ref{sec:fac}
we review the basics of factorization which will be essential for our
arguments. We select the mesons for which factorizing contributions in
decays are suppressed in Sect.~\ref{sec:mesons}. In the following
section we identify which flavor structure a decay must have in order
for this suppression to apply, and take a closer look at specific
issues in the different channels. Sect.~\ref{sec:break} discusses how
suppression can be circumvented by different nonfactorizing
mechanisms. Some of these can be treated within QCD factorization and
will be investigated in Sect.~\ref{sec:bbns}. We estimate branching
ratios of suppressed decays into a heavy and a light meson in the
subsequent section, before concluding in Sect.~\ref{sec:sum}. Some
numerical estimates concerning meson distribution amplitudes, which we
need in our paper, are given in an Appendix.

%%%%%%%%%%%%%%%%%%%%%%%%%%%%%%%%%%%%%%%%%%%%%%%%%%%%%%%%%%%%%%%%%%%
\section{Matrix elements for hadronic two-body decays}
\label{sec:fac}
%%%%%%%%%%%%%%%%%%%%%%%%%%%%%%%%%%%%%%%%%%%%%%%%%%%%%%%%%%%%%%%%%%%

We start by briefly recalling the low-energy effective Hamiltonian and
some basics of the factorization approach.  Hadronic two-body
$b$-decays are described by the effective weak Hamiltonian 
\begin{eqnarray}
\label{eq:heff}
{\cal{H}}_{\mathrm{eff}}=
\frac{G_F}{\sqrt{2}} \left[ \, \sum_{j,k=u,c} 
V_{jb}^{\phantom{*}} V^*_{kd} \,
\Big( C_1^{\phantom{j}} O_1^{jk} + C_2^{\phantom{j}} O_2^{jk} \Big) -
V_{tb}^{\phantom{*}} V^*_{td} \sum_i C_i O_i \,\right] 
+ \{d \rightarrow s\} + \mathrm{h.c.} \, ,
\end{eqnarray}
where $V$ denotes the CKM matrix. The operators $O_{1,2}^{jk}$ result
from tree level $W$ exchange and in the case $i=k=u$ read
\begin{eqnarray}
O_1^{uu}&=& 
\bar{u}_\alpha \gamma^\mu (1-\gamma_5) b_\alpha\:
\bar{d}_\beta \gamma_\mu (1-\gamma_5) u_\beta ,
\nonumber \\
O_2^{uu}&=& 
\bar{u}_\alpha \gamma^\mu (1-\gamma_5) b_\beta\:
\bar{d}_\beta \gamma_\mu (1-\gamma_5) u_\alpha .
\end{eqnarray}
Here $\alpha$ and $\beta$ are color indices, and it is understood that
all fields are taken at space-time argument zero. The remaining
operators in ${\cal{H}}_{\mathrm{eff}}$ are so-called penguins. For a
detailed discussion of the operators $O_i$ and the Wilson coefficients
$C_i$ we refer to \cite{Buchalla:1996vs}.

In naive factorization, the matrix element $\langle Y X|\,
{\cal{H}}_{\mathrm{eff}} \,| B\rangle$ is written as a product of
matrix elements of quark currents between $B$ and $Y$, and between the
vacuum and $X$. Only the color-singlet piece of each current is
retained, while the color-octet piece is neglected. This leads to
replacing ${\cal{H}}_{\mathrm{eff}}$ by the effective transition
operator ${\cal T}$, whose tree level operators read
\begin{eqnarray}
{\cal T}^{(1,2)} &=&
\frac{G_F}{\sqrt{2}}\, V^{\phantom{*}}_{ub} V_{ud}^*\, \Big[\,
a_1\; \bar{u} \gamma^\mu (1-\gamma_5) b 
    \otimes \bar{d} \gamma_\mu (1-\gamma_5) u
\nonumber \\
&& \hspace{4em} {}+
a_2\; \bar{d} \gamma^\mu (1-\gamma_5) b 
    \otimes \bar{u} \gamma_\mu (1-\gamma_5) u \,\Big] 
\label{t-12}
\end{eqnarray}
for $i=k=u$. Here the notation $\otimes$ indicates that the matrix
elements are to be taken in factorized form as described above. The
new coefficients $a_1=C_1 + C_2/3$ and $a_2=C_2 + C_1/3$ have been
obtained from projecting on color-singlet currents, and are commonly
referred to as color allowed and color suppressed,
respectively. Numerically, $a_1$ is close to 1 and $a_2$ of order 0.1
at a renormalization scale $\mu=m_b$. Notice that the Fierz transform
performed in the second term of Eq.~(\ref{t-12}) has left the $(V-A)
\times (V-A)$ structure invariant, where $V$ and $A$ respectively
denote the vector and axial vector current.  The situation is
analogous for those penguin operators which again involve $(V-A)
\times (V-A)$ currents, for explicit formulae see
e.g.~\cite{Beneke:2001ev}.

Important for us will be the strong penguins with $(V-A) \times (V+A)$
structure (the electroweak ones with similar Dirac structure are
numerically less important in the Standard Model). They are
\begin{eqnarray}
O_5&=& 
\bar{d}_\alpha \gamma^\mu (1-\gamma_5) b_\alpha\:
\sum_q \bar{q}_\beta \gamma_\mu (1+\gamma_5) q_\beta ,
\nonumber \\
O_6&=& 
\bar{d}_\alpha \gamma^\mu (1-\gamma_5) b_\beta\:
\sum_q \bar{q}_\beta \gamma_\mu (1+\gamma_5) q_\alpha ,
\end{eqnarray}
with a sum over $q=u,d,s,c,b$, and in naive factorization lead to
\begin{eqnarray}
{\cal T}^{(5,6)} &=&
- \frac{G_F}{\sqrt{2}}\, V^{\phantom{*}}_{tb} V_{td}^*\, \Big[\,
a_5\; \sum_q \bar{d} \gamma^\mu (1-\gamma_5) b 
    \otimes \bar{q} \gamma_\mu (1+\gamma_5) q
\nonumber \\
&& \hspace{4.5em} {}+
a_6\; \sum_q\, (-2)\, \bar{q} (1-\gamma_5) b 
    \otimes \bar{d} (1+\gamma_5) q \;\Big] .
\label{t-56}
\end{eqnarray}
The structure $(P-S) \times (P+S)$ involving the scalar and
pseudoscalar currents $S$ and $P$ has emerged from the Fierz transform
in the term with $a_6=C_6+C_5/3$, whose value is about $-0.03$ at
$\mu=m_b$. The corresponding operator provides one possibility to
circumvent the suppression mechanisms we will discuss shortly, and we
will often refer to it as scalar penguin.

%%%%%%%%%%%%%%%%%%%%%%%%%%%%%%%%%%%%%%%%%%%%%%%%%%%%%%%%%%%%%%%%%%%
\section{Meson candidate selection}
\label{sec:mesons}
%%%%%%%%%%%%%%%%%%%%%%%%%%%%%%%%%%%%%%%%%%%%%%%%%%%%%%%%%%%%%%%%%%%

Let us now specify our mechanisms to suppress decay amplitudes in
naive factorization and see to which final states they apply.

%---------------------------------------------------------------------

\subsection{Suppression mechanisms}
\label{sec:suppress}

There are several reasons why the coupling of a meson to the local
currents of the effective weak Hamiltonian can be suppressed. Clearly,
a meson with spin $J=2$ or larger has no matrix element with either of
the currents $S$, $P$, $V$, $A$, and in naive factorization cannot be
produced as a meson ejected by the effective weak current. {}From
Table~\ref{tab:friends} we see that examples for such mesons are the
$a_2$, $\pi_2$, $\rho_3$, and $K^{*}_2$ in the light quark
sector. Heavy tensor mesons are the $D_2^*$ and $D_{sJ}$, and there is
a tensor charmonium state, $\chi_{2c}$.

\begin{table}
\begin{center}
  \begin{tabular}{lccccccccccc} \hline\hline
$X$ &         $a_0$    & $b_1$    & $\pi$    & $a_2$    & $a_0$    &
              $\pi_2$  & $\rho_3$ & $\pi$    & $a_4$    &
              $\chi_{c0}$ & $\chi_{c2}$ \\
$m_X$ [MeV] & ~985     & 1230     & 1300     & 1318     & 1474     &
              1670     & 1691     & 1801     & 2014     &
              3415        & 3556  \\
$J^{PC}$ &    $0^{++}$ & $1^{+-}$ & $0^{-+}$ & $2^{++}$ & $0^{++}$ &
              $2^{-+}$ & $3^{--}$ & $0^{-+}$ & $4^{++}$ &
              $0^{++}$    & $2^{++}$  \\
  \hline \hline
  \end{tabular} \\[\baselineskip]
  \begin{tabular}{lcccccccc} \hline\hline
$X$ &         $K_0^*$ & $K_2^*$ & $K_2$   & $K_3^*$ & $K_2$   &
              $K_4^*$ & $D_2^*$ & $D_{sJ}$ \\
$m_X$ [MeV] & 1412    & 1426    & 1773    & 1776    & 1816    &
              2045    & 2459    & 2573 \\
$J^{P}$ &     $0^{+}$ & $2^{+}$ & $2^{-}$ & $3^{-}$ & $2^{-}$ &
              $4^{+}$ & $2^+$   & $(2^+)$ \\
  \hline \hline
  \end{tabular}
\caption{\label{tab:friends}Mesons for which one of the suppression
mechanisms discussed in Sect.~\ref{sec:suppress} is relevant.
Masses are taken from the Review of Particle
Data~\cite{Groom:2000in}. We do not list those mesons that are omitted
in the meson summary table. The spin-parity assignment for the
$D_{sJ}(2537)$ is not certain yet.}
\end{center}
\end{table}

Let us now turn to mesons with $J=0,1$ whose production from the weak
current in naive factorization is forbidden or suppressed because
their coupling to $V$ and $A$ is zero or small. We define the decay
constants of the negatively charged mesons with isospin $I=1$
as
\begin{eqnarray}
\vev{\bar{d}(0)\, \gamma^{\mu}\, u(0)}{S(q)} &=& -i f_S\, q^\mu ,
\label{eq:a0} \\
\vev{\bar{d}(0)\, \gamma^{\mu} \gamma_5\, u(0)}{P(q)} &=& 
       -i f_P\, q^\mu ,
\label{eq:pi} \\
\vev{\bar{d}(0)\, \gamma^{\mu}\, u(0)}{V(q,\epsilon)} &=& 
       -i f_V m_V \epsilon^{\mu} ,
\\
\vev{\bar{d}(0)\, \gamma^{\mu} \gamma_5\, u(0)}{A(q,\epsilon))} &=& 
       -i f_A m_A \epsilon^{\mu} ,
\label{eq:b1}
\end{eqnarray}
for scalar, pseudoscalar, vector, and axial mesons, respectively.
Here $q^{\mu}$ denotes the meson momentum and, if applicable,
$\epsilon^{\mu}$ its polarization vector.  The choice of vector or
axial quark current on the right-hand sides of these definitions is
dictated by the parity of the meson.  For the corresponding neutral
mesons, the flavor structure of the current is $(\bar{u} u - \bar{d}
d)/ \sqrt{2}$ instead of $\bar{d} u$, and for mesons with different
quark content one has to take $\bar{s}d$, $\bar{s}u$, $\bar{c}c$, etc.

Because of charge conjugation invariance, the decay constants for the
neutral $a_0$ and $b_1$ mesons and for the $\chi_{0c}$ must be zero.
In the isospin limit, the decay constants of the charged $a_0$ and
$b_1$ must thus vanish, too, so that $f_{a0}$ and $f_{b1}$ defined in
Eqs.~(\ref{eq:a0}), (\ref{eq:b1}) are small, of order $m_d - m_u$. For
the $a_0$ mesons, this can explicitly be seen by taking the divergence
of Eq.~(\ref{eq:a0}), which by virtue of the equations of motion gives
\begin{eqnarray}
m_{a_0}^2\,   f_{a_0}^{\phantom{a_0}}
 &=& i (m_d-m_u)\, \vev{\bar{d}(0) u(0)}{a_0} .
  \label{eq:f_a0}
\end{eqnarray}
For the charged $K^*_0$ the analogous relation reads
\begin{eqnarray}
m_{K^*_0}^2\, f_{K^*_0}^{\phantom{K^*_0}}
 &=& i (m_s-m_u)\, \vev{\bar{s}(0) u(0)}{K^*_0} ,
  \label{eq:f_K0}
\end{eqnarray}
which becomes zero in the flavor SU(3) limit and indicates that
$f_{K^*_0}$ should be suppressed.

It is instructive to compare Eqs.~(\ref{eq:f_a0}) and (\ref{eq:f_K0})
with their analogs for the pseudoscalars,
\begin{eqnarray}
m_{\pi}^2\, f_{\pi}^{\phantom{\pi}} &=& i (m_d+m_u)\, 
            \vev{\bar{d}(0) \gamma_5 u(0)}{\pi} ,
  \label{eq:f_pi} \\
m_{K}^2\,   f_{K}^{\phantom{K}}     &=& i (m_s+m_u)\, 
            \vev{\bar{s}(0) \gamma_5 u(0)}{K} .
  \label{eq:f_K}
\end{eqnarray}
Because the axial current is not conserved, these decay constants do
not vanish in the isospin or SU(3) limit. Moreover, they do \emph{not}
vanish in the chiral limit, $m_u=m_d=m_s=0$, for the light pion and
kaon, since these mesons are Goldstone bosons and become massless in
the same limit. Numerically, $f_\pi$ and $f_K$ are in fact not small
and of the same order of magnitude as for instance $f_\rho$ and
$f_{K^*}$. The decay constant for the heavy $\pi(1300)$, however,
\emph{does} become zero in the chiral limit, and its actual value is
small due to chiral suppression.

We remark that the spin-zero mesons whose coupling to the $V$ and $A$
currents is small or zero for one of the above reasons can still
couple to the $S$ or $P$ currents appearing in penguin operators of
the effective Hamiltonian, as discussed in Sect.~\ref{sec:fac}. This
does however not hold for the $b_1$, which has no matrix elements with
$S$ or $P$.

\subsection{Decay constants}

The decay constants of the $a_0(980)$, $a_0(1450)$, $K^*_0$,
$\pi(1300)$, and the $b_1$ are poorly known at present. Using finite
energy sum rules, Maltman \cite{Maltman:1999jn}
obtained\footnote{Maltman defines the $a_0$ decay constants with an
extra factor $(m_s-m_u)/(m_d-m_u)$. To convert them into our
convention, we take the quark masses in
Eq.~(\protect\ref{quark-masses}) below.}
\begin{equation}
f_{a0(980)}  = 1.1  \mbox{~MeV} , \hspace{2em}
f_{a0(1450)} = 0.7 \mbox{~MeV} , \hspace{2em}
f_{K^*_0} = 42 \mbox{~MeV} ,
\label{malt}
\end{equation}
consistent with the ranges estimated by Narison \cite{Narison:1989aq}
\begin{equation}
f_{a0(980)} = 0.7 \mbox{~to~} 2.5 \mbox{~MeV} , \hspace{2em}
f_{K^*_0} = 33 \mbox{~to~} 46 \mbox{~MeV} .
\label{nar}
\end{equation}
For the heavy pion, the theoretical estimates in \cite{Volkov:1997br}
provide a range
\begin{equation}
f_{\pi(1300)} = 0.5 \mbox{~to~} 7.2 \mbox{~MeV} .
\label{heavy-pi}
\end{equation}
Comparing these values to
\begin{equation}
f_{\pi} = 131 \mbox{~MeV} , \hspace{2em} f_{K} = 160 \mbox{~MeV} ,
\label{light-pi}
\end{equation}
we find that the suppression patterns discussed in the previous
subsection are indeed seen numerically, with the decay constants for
the $a_0$ mesons smaller than those for $\pi(1300)$ because of the
relative signs between the quark masses in Eqs.~(\ref{eq:f_a0}) and
(\ref{eq:f_pi}). We also see that $f_{K^*_0}$ is suppressed relative
to $f_{K}$, but not as strongly as $f_{a_0}$ compared with $f_{\pi}$,
because SU(3) symmetry breaking is rather strong for the quark masses.
Here we have implicitly assumed that the (pseudo)scalar matrix
elements on the right-hand sides of Eqs.~(\ref{eq:f_a0}) to
(\ref{eq:f_K}) are not anomalously small or large. Indeed, taking
from~\cite{Narison:2000uj}\,\footnote{We have taken the average values
given in Table~6 and evolved them down using Eq.~(20) in that
reference.}
\begin{equation}
m_u = 4.8\mbox{~MeV}, \hspace{2em} m_d = 8.7\mbox{~MeV}, \hspace{2em}
m_s = 164\mbox{~MeV}
\label{quark-masses}
\end{equation}
for the $\overline{\mbox{MS}}$ quark masses at $\mu=1$~GeV, we find
with the values~(\ref{heavy-pi}) and (\ref{light-pi})
\begin{eqnarray}
i\, \vev{\bar{d} \gamma_5 u}{\pi} &\approx& 0.19 \mbox{~GeV}^2 ,
\nonumber \\
i\, \vev{\bar{d} \gamma_5 u}{\pi(1300)} &\approx& 
0.06 \mbox{~to~} 0.90 \mbox{~GeV}^2 ,
\nonumber \\
i\, \vev{\bar{s} \gamma_5 u}{K} &\approx& 0.23 \mbox{~GeV}^2 ,
\label{pseudo-constants}
\end{eqnarray}
and with (\ref{malt}) and (\ref{nar})
\begin{eqnarray}
i\, \vev{\bar{d} u}{a_0(980)} &\approx& 
0.17 \mbox{~to~} 0.62 \mbox{~GeV}^2 , \nonumber \\
i\, \vev{\bar{d} u}{a_0(1450)} &\approx& 
0.39 \mbox{~GeV}^2 , \nonumber \\
i\, \vev{\bar{s} u}{K^*_0(1430)} &\approx& 
0.42 \mbox{~to~} 0.58 \mbox{~GeV}^2 .
\label{scalar-constants}
\end{eqnarray}
Despite a certain spread these values are remarkably close to each
other, given that the corresponding squared meson masses vary by more
than two orders of magnitude. We note that Chernyak
\cite{Chernyak:2001hs} has recently estimated $f_{K^*_0} = (70\pm
10)$~MeV. We consider this to be rather high as it is far away from
the range (\ref{nar}) obtained in other studies. Also, the
corresponding value for $\vev{\bar{s} u}{K^*_0(1430)}$ would
correspond to a quite strong SU(3) breaking for the scalar matrix
elements.

We wish to emphasize at this point that the decay constants for the
$a_0(980)$, $a_0(1450)$, $K^*_0$, $\pi(1300)$, and the $b_1$ can be
measured very cleanly in $\tau$ decays. In fact, from the bound on the
branching ratio ${\cal{B}}(\tau \to \pi(1300) \nu_\tau) < 1 \cdot
10^{-4}$ in \cite{Groom:2000in} we infer $f_{\pi(1300)} < 8.4$~MeV,
which is not far from the upper end of the theory estimates
(\ref{heavy-pi}). The decay constants in Eq.~(\ref{malt}) correspond
to branching ratios of
\begin{eqnarray}
{\cal{B}}(\tau \to a_0(980)\, \nu_\tau) 
   & \simeq & 3.8\cdot 10^{-6} , \nonumber \\
{\cal{B}}(\tau \to a_0(1450)\, \nu_\tau)
   & \simeq & 3.7\cdot 10^{-7} , \nonumber \\
{\cal{B}}(\tau \to K^*_0(1430)\, \nu_\tau)
   & \simeq & 7.7\cdot 10^{-5} .
\label{eq:taudecays}
\end{eqnarray}
These estimates are rather encouraging, given that at the $B$
factories one expects to have about $3\cdot 10^7$ $\tau$ pairs with
30~fb$^{-1}$ \cite{Harrison:1998yr}, and that the potential of
$\tau$-charm factories would be even higher. A measurement of the
decay constants for the above mesons would greatly reduce the
uncertainties in the predictions we will give in
Sect.~\ref{sec:branching}. It could also provide valuable information
on the nature in particular of the $a_0(980)$ and $a_0(1450)$, only
one of which can be a member of the conventional $q\bar{q}$ meson
nonett.

%---------------------------------------------------------------------

\subsection{Kinematics}
\label{sec:kinematics}

\begin{figure}
\begin{center}
\leavevmode
\epsfxsize=0.55\textwidth
\epsffile{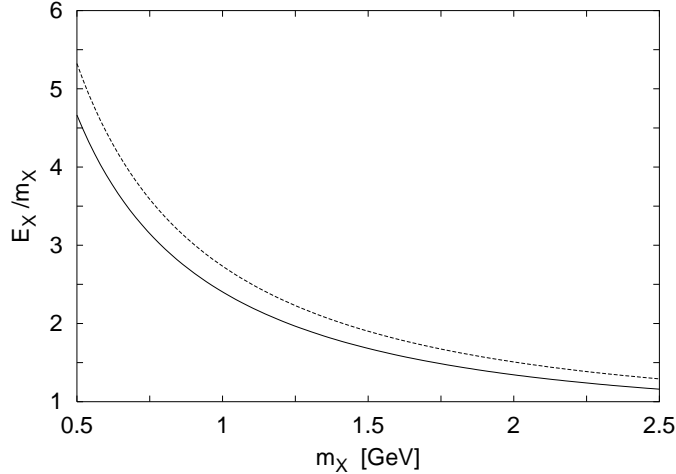}
\end{center}
\caption[]{Values of $E_X/m_X$ as a function of $m_X$ for the decays
$B\to D X$ (lower curve) and $B\to \pi X$ (upper curve).}
\label{fig:kin}
\end{figure}

The color transparency argument \cite{Bjorken:1989kk} for
factorization of decays $B\to Y X$ requires the meson $X$ emitted from
the weak current to be fast. More quantitatively, its time dilation
factor $E_X /m_X$ should be large, where $E_X = (m_B^2 - m_Y^2 +
m_X^2) /(2 m_B)$ is the energy of $X$ in the $B$ meson rest frame. We
show the values of $E_X /m_X$ for $Y=D$ and $Y=\pi$ in
Fig.~\ref{fig:kin}. The corresponding curves for $B\to D^* X$ and $B_s
\to D_s^{(*)} X$ are very close to the one for $B\to D X$, and the
ones for $B\to \rho X$, $B_s\to K X$, and $B_s\to K^*(892)\, X$ are
practically the same as for $B\to \pi X$. Only if $X$ is a pion does
one have a very large $E_X /m_X$, namely $E_\pi/m_\pi=16.5$ for $B\to
D \pi$ and $E_\pi/m_\pi=19$ for $B\to \pi \pi$. For $m_X$ above 1~GeV,
relevant for the mesons in Table~\ref{tab:friends}, this ratio
decreases rather gently from moderate values down to little above
1. Even so, the mass range of our candidate mesons seems sufficiently
large so that a study of the corresponding decay channels could
provide valuable clues to whether corrections to factorization
significantly depend on $E_X /m_X$, or more generally, on the mass
$m_X$. We remark in passing that even the lowest value of $E_X/m_X$ in
Fig.~\ref{fig:kin} corresponds to a velocity $\beta_X=0.5$ and a
recoil momentum of $p_X=1.5$~GeV in the $B$ rest frame, indicating
that $X$ is still relativistic.

%---------------------------------------------------------------------

\subsection{Resonance decays and the continuum}
\label{sec:continuum}

Clearly, the measurement of rare $B$ decays involving higher mass
resonances presents an experimental challenge. An experimental
analysis will be easier if the meson $X$ in question has a decay
channel with sizeable branching fraction that is not also accessible
to mesons with nearby mass whose production is not suppressed. To give
some examples, the decays of $a_0(980)$, $a_2(1320)$, and $a_0(1450)$
into $\pi\eta$ appear rather clean in this respect, as do the modes
$b_1(1235)\to \omega\pi$ and $\pi_2(1670)\to f_2(1270) \pi$. On the
other hand, the decays of $a_2(1320)$, $\pi(1300)$, and $\pi_2(1670)$
into $\rho\pi$ are more problematic because of background from the
rather broad $a_1(1260)$. The same holds true for the decays of the
$K_0^*(1430)$ and $K_2^*(1430)$ into $K\pi$ because of background from
the $K^*(1410)$. In such cases a partial wave analysis of the decay
products will probably be necessary in order to constrain the decays
into the mesons with spin $J=0$ or $J=2$.

We emphasize at this point that mesons $X$ with a rather large decay
width do not present as serious a problem in our context as in other
studies. The physical arguments leading to a suppression of their
production within the factorization mechanism (as the arguments for
factorization itself) do in fact not depend on $X$ being a narrow
resonance. Our arguments in Sect.~\ref{sec:suppress} were at the level
of current matrix elements and go through in complete analogy if, for
instance, the $|\pi(1300)\rangle$ state in Eq.~(\ref{eq:f_pi}) is
replaced with $|\rho\pi\rangle$ in appropriate partial waves that have
definite quantum numbers $J^P=0^-$. Moreover, the main idea here is to
use the branching ratios of suppressed decays as quantitative
estimates for the size of corrections to factorization, and to study
their pattern by comparing different decay channels. An uncertainty on
the branching ratio of $B\to Y X$ due to the line shape of $X$ is
therefore less severe as in channels which are allowed by
factorization, and where factorization tests need branching ratios to
a much higher precision.

One could in fact also perform the studies we propose here not with
particular meson resonances but with continuum states, similarly to a
recent test of factorization by Ligeti et al.~\cite{Ligeti:2001dk}.
Our main reason to concentrate on resonances $X$ here is that, by
definition, their production should be enhanced with respect to
continuum states with the same quantum numbers, which is important
since we are looking for decays with small branching ratios from the
start.

%%%%%%%%%%%%%%%%%%%%%%%%%%%%%%%%%%%%%%%%%%%%%%%%%%%%%%%%%%%%%%%%%%%
\section{Decay mode selection}
\label{sec:flavor}
%%%%%%%%%%%%%%%%%%%%%%%%%%%%%%%%%%%%%%%%%%%%%%%%%%%%%%%%%%%%%%%%%%%

We are looking for exclusive decays $B \to Y X$, where the meson $X$
must be emitted from the weak decay vertex and cannot pick up the
spectator quark.  Only then will the factorizing contribution to the
decay be suppressed for mesons $X$ with small or vanishing decay
constant or with spin $J\ge 2$. This puts requirements on the flavor
structure of the decay, which we now discuss. One may avoid these
requirements by studying decays where \emph{both} final state mesons
are taken from Table~\ref{tab:friends}, such as $\bar{B}^0\to b_1^+
b_1^-$ or $\bar{B}^0\to \pi_2^+ a_2^-$. For one of the mesons the
appropriate suppression mechanism will then always be at work.

For definiteness we consider in the following the case where the $B$
meson contains a $b$ and not a $\bar{b}$.  One requirement now is that
the flavors of the spectator antiquark and of the antiquark emitted
from the $b$ decay must not be the same, otherwise $X$ can pick up
either of them. We thus cannot use decays such as $B^-\to D^0 a_0^-$,
whose flavor structure reads $\bar{u}b\to \bar{u}(c\bar{u}d)$, where
the brackets indicate the quarks originating from the electroweak
vertex.

\begin{figure}
\begin{center}
\leavevmode
\epsfxsize=0.9\textwidth
\epsffile{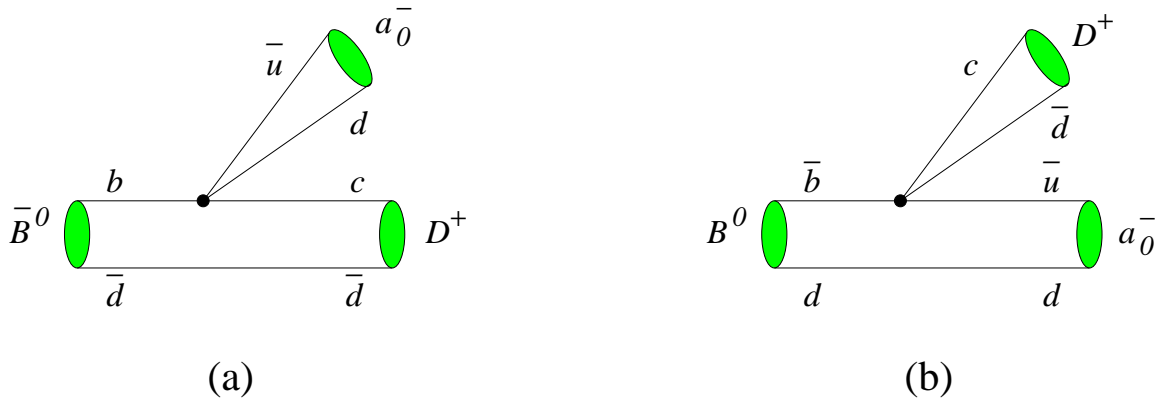}
\end{center}
\caption[]{Tree level diagrams for (a) $\bar{B}^0 \to D^{+} a_0^-$
and (b) $B^0 \to D^{+} a_0^-$.}
\label{fig:tree}
\end{figure}

A second requirement is due to imperfect knowledge of the initial
state. Whereas the decay $\bar{B}^0\to D^+ a_0^-$ with flavor
structure $\bar{d}b\to \bar{d}(c\bar{u}d)$ satisfies our requirement,
the same final state can be reached in the decay of a $B^0$, where the
$a_0^-$ contains the spectator. The corresponding diagrams are shown
in Fig.~\ref{fig:tree}. This $C\!P$ conjugated background can in
principle be removed by flavor tagging, which puts of course stronger
demands on the experiment. In the example just given, the amplitude
from $B^0$ decay is suppressed by $\lambda^2$ relative to the direct
decay, where $\lambda\approx 0.22$ is the Wolfenstein parameter in the
CKM matrix. We will give a more quantitative estimate of this
background in Sect.~\ref{sec:background}. In other cases however,
e.g.\ for $\bar{B}_s\to D_s^+ K_0^{*-}$ versus $B_s\to D_s^+
K_0^{*-}$, both signal and background amplitudes are of order
$\lambda^3$. We will not consider such modes in the following, and
list in Tables~\ref{tab:flavor} and \ref{tab:flavor-two} the flavor
structure of decays satisfying the following conditions:
\begin{enumerate}
\item It is ensured that the meson selected from
Table~\ref{tab:friends} cannot pick up the spectator antiquark in the
$B$ meson.
\item A $C\!P$ conjugated background that violates condition 1 either
does not exist or is CKM suppressed. It turns out that the only case
we retain that has such a background is the one mentioned above,
listed in the first row of Table~\ref{tab:flavor}.
\end{enumerate}
Let us now consider the different decay categories separately.

\begin{table}
\begin{center}
\renewcommand{\arraystretch}{1.2}
\begin{tabular}{lcccccc} \hline \hline
example decay & \multicolumn{3}{c}{factorizing contribution} & 
        \multicolumn{3}{c}{annihilation} \\
 & quark level & tree & peng. & topology & tree & peng. \\ \hline
$\bar{B}^0 \to D^{+} a_0^-$   & 
  $\bar{d}b\to \bar{d}(c\bar{u}d)$ &
  $\lambda^2$ & & a & $\lambda^2$ & \\
$\phantom{\bar{B}^0}\to D^{+} K_0^{*-}$ &
  $\phantom{\bar{d}b}\to \bar{d}(c\bar{u}s)$ & 
  $\lambda^3$ & & & \\
$\bar{B}_s \to D_s^{+} a_0^-$        & 
  $\bar{s}b\to \bar{s}(c\bar{u}d)$ &
  $\lambda^2$ & & & \\ 
\hline
$\bar{B}^0 \to \pi^+ D_{sJ}^{-}$ & 
  $\bar{d}b\to \bar{d}(u\bar{c}s)$ & $\lambda^3$ \\
$B^- \to \pi^0 D_2^{*-}$ &        
  $\bar{u}b\to \bar{u}(u\bar{c}d)$ & 
  $\lambda^4$ & & a & $\lambda^4$ & \\
$\phantom{B^-} \to \eta D_2^{*-}$ &        
  $\phantom{\bar{u}b}\to \bar{u}(u\bar{c}d)$ & 
  $\lambda^4$ & & a, c & $\lambda^4$ & \\
$\phantom{B^-} \to \pi^0 D_{sJ}^{-}$ &          
  $\phantom{\bar{u}b}\to \bar{u}(u\bar{c}s)$ & 
  $\lambda^3$ & & & & \\
$\phantom{B^-} \to \eta D_{sJ}^{-}$ &          
  $\phantom{\bar{u}b}\to \bar{u}(u\bar{c}s)$ & 
  $\lambda^3$ & & a, c & $\lambda^3$ & \\
$\bar{B}_s \to K^+ D_2^{*-}$ &      
  $\bar{s}b\to \bar{s}(u\bar{c}d)$ & $\lambda^4$ \\ 
\hline
$\bar{B}^0 \to D^{+} D_{sJ}^{-}$ & 
  $\bar{d}b\to \bar{d}(c\bar{c}s)$ &
  $\lambda^2$ & $\lambda^2$ & a & & $\lambda^2$ \\
$B^- \to D^{0} D_2^{*-}$ &          
  $\bar{u}b\to \bar{u}(c\bar{c}d)$ & 
  $\lambda^3$ & $\lambda^3$ & a & $\lambda^3$ & $\lambda^3$ \\
$\phantom{B^-}\to D^{0} D_{sJ}^{-}$ &          
  $\phantom{\bar{u}b}\to \bar{u}(c\bar{c}s)$ & 
  $\lambda^2$ & $\lambda^2$ & a & $\lambda^4$ & $\lambda^2$ \\
$\bar{B}_s \to D_s^{+} D_2^{*-}$ &      
  $\bar{s}b\to \bar{s}(c\bar{c}d)$ & 
  $\lambda^3$ & $\lambda^3$ & a & & $\lambda^3$ \\ 
\hline
$\bar{B}^0 \to \pi^+ K_2^{*-}$ & 
  $\bar{d}b\to \bar{d}(u\bar{u}s)$ & 
  $\lambda^4$ & $\lambda^2$ & a & & $\lambda^2$ \\
$B^-\to \pi^- \bar{K}_2^{*0}$ & 
  $\bar{u}b\to \bar{u}(d\bar{d}s)$ &
   & $\lambda^2$ & a & $\lambda^4$ & $\lambda^2$ \\
$\phantom{B^-}\to K^- K_2^{*0}$ & 
  $\phantom{\bar{u}b}\to \bar{u}(s\bar{s}d)$ &
   & $\lambda^3$ & a & $\lambda^3$ & $\lambda^3$ \\
$\bar{B}_s\to K^+ a_2^-$       & 
  $\bar{s}b\to \bar{s}(u\bar{u}d)$ & 
  $\lambda^3$ & $\lambda^3$ & a & & $\lambda^3$ \\ 
\hline \hline
\end{tabular}
\end{center}
\caption{\label{tab:flavor} Color allowed decay modes $B\to Y X$
satisfying the criteria specified in the text. $X$ is the meson
emitted from the weak decay vertex. Any meson can be replaced by
another with the same flavor structure and isospin, e.g, $D$ by $D^*$,
$\pi$ by $\rho$, and the meson $X$ by an appropriate candidate from
Table~\protect\ref{tab:friends}. The second column gives the quark
composition of the final state, with the quarks originating from the
$b$ decay enclosed in brackets.  In the third and fourth columns we
give the power of the Wolfenstein parameter $\lambda$ in the decay
amplitude for tree level $W$ exchange and penguin operators,
respectively. Annihilation contributions are listed with their
topology as shown in Fig.~\ref{fig:anni} and the power of the
Wolfenstein parameter for tree level and penguin operators.}
\end{table}

\begin{table}[tb]
\begin{center}
\renewcommand{\arraystretch}{1.2}
\begin{tabular}{lcccccc} \hline \hline
example decay & \multicolumn{3}{c}{factorizing contribution} & 
        \multicolumn{3}{c}{annihilation} \\
 & quark level & tree & peng. & topology & tree & peng. \\ \hline
$\bar{B}^0 \to \pi^0 D_2^{*0}$ & 
  $\bar{d}b \to \bar{d}(d\bar{u}c)$ & 
  $\lambda^2$ & & a & $\lambda^2$ & \\
$\phantom{\bar{B}^0} \to \eta\, D_2^{*0}$ & 
  $\phantom{\bar{d}b} \to \bar{d}(d\bar{u}c)$ & 
  $\lambda^2$ & & a, c & $\lambda^2$ & \\
$B^- \to K^- D_2^{*0}$ &
  $\bar{u}b \to \bar{u}(s\bar{u}c)$ & $\lambda^3$ \\
$\bar{B}_s \to K^0 D_2^{*0}$ &
  $\bar{s}b \to \bar{s}(d\bar{u}c)$ & $\lambda^2$\\
$\phantom{\bar{B}_s} \to \eta\, D_2^{*0}$ &
  $\phantom{\bar{s}b} \to \bar{s}(s\bar{u}c)$ &
  $\lambda^3$ & & a, c & $\lambda^3$ & \\
\hline
$\bar{B}^0\to \bar{K}^0 \chi_{c0}$ & 
  $\bar{d}b\to \bar{d}(s\bar{c}c)$ &
  $\lambda^2$ & $\lambda^2$ & b & & $\lambda^2$ \\
$\phantom{\bar{B}^0}\to \pi^0 \chi_{c0}$ & 
  $\phantom{\bar{d}b}\to \bar{d}(d\bar{c}c)$ &
  $\lambda^3$ & $\lambda^3$ & b & $\lambda^3$ & $\lambda^3$ \\
$\phantom{\bar{B}^0}\to \eta\, \chi_{c0}$ & 
  $\phantom{\bar{d}b}\to \bar{d}(d\bar{c}c)$ &
  $\lambda^3$ & $\lambda^3$ & b, c & $\lambda^3$ & $\lambda^3$ \\
$B^-\to K^- \chi_{c0}$ & 
  $\bar{u}b\to \bar{u}(s\bar{c}c)$ &
  $\lambda^2$ & $\lambda^2$ & b & $\lambda^4$ & $\lambda^2$ \\
$\phantom{B^-}\to \pi^- \chi_{c0}$ & 
  $\phantom{\bar{u}b}\to \bar{u}(d\bar{c}c)$ &
  $\lambda^3$ & $\lambda^3$ & b & $\lambda^3$ & $\lambda^3$ \\
$\bar{B}_s\to \eta\, \chi_{c0}$ & 
  $\bar{s}b\to \bar{s}(s\bar{c}c)$ &
  $\lambda^2$ & $\lambda^2$ & b, c & $\lambda^2$ & $\lambda^2$ \\
$\phantom{\bar{B}_s}\to K^0 \chi_{c0}$ & 
  $\phantom{\bar{s}b}\to \bar{s}(d\bar{c}c)$ &
  $\lambda^3$ & $\lambda^3$ & b & & $\lambda^3$ \\
\hline
$\bar{B}_s \to K^0\, a_2^0$ & 
  $\bar{s}b\to \bar{s}(d\bar{u}u)$ &
  $\lambda^3$ & $\lambda^3$ & a & & $\lambda^3$ \\
$\phantom{\bar{B}_s}\to \eta\; a_2^0$ & 
  $\phantom{\bar{s}b}\to \bar{s}(s\bar{u}u)$ &
  $\lambda^4$ & $\lambda^2$ & a, c & $\lambda^4$ & $\lambda^2$ \\
\hline \hline
\end{tabular}
\end{center}
  \caption{\label{tab:flavor-two} As Table~\protect\ref{tab:flavor}
but for decays where the tree level contribution is color
suppressed. We remark that the decay $\bar{B}_s \to K^0\, a_2^0$ has a
color allowed penguin contribution.}
\end{table}

%---------------------------------------------------------------------
\subsection{Decays with one or two heavy mesons}
\label{sec:light-heavy}
%---------------------------------------------------------------------

Decays into heavy-light final states with open charm are the simplest
from the point of view of their electroweak structure, since only the
$W$ exchange operators $O_1$ and $O_2$ of the effective Hamiltonian
contribute. For color allowed decays $B\to D X$ with $X=\pi, \rho,
a_1$, naive factorization is in rather good agreement with
data~\cite{Neubert:1997uc,Luo:2001mc}.

For color allowed decays where the heavy meson is the emission
particle, the color transparency argument does not hold, but arguments
based on the large $N_c$ limit do. Relevant meson candidates here are
the $D_2^*$ and the $D_{sJ}$. Comparing the size of nonfactorizing
contributions for the two types of channels with the suppression
mechanisms discussed here might thus shed light on the question which
type of mechanism is more relevant to ensure factorization. One may
also use decays into two charmed mesons to address the same
question. We remark in this context that in recent study, Luo and
Rosner found factorization to work reasonably well for $\bar{B}^0\to
D^{(*)+} D_s^{(*)-}$ within present errors~\cite{Luo:2001mc}.

Notice that for color suppressed channels such as $\bar{B}^0\to \pi^0
D_2^{*0}$ naive factorization is neither backed up by color
transparency nor by $1/N_c$ arguments. The decays into a $D_2^{*0}$ in
Table~\ref{tab:flavor-two} will thus show whether the factorization
concept still applies here.

Let us finally consider $B$ decays into charmonium. Naive
factorization has notorious problems with these channels
\cite{Neubert:1997uc}, compounded by the fact that the coefficient
$a_2$ is extremely dependent on the factorization scale $\mu$. A
comparison of the decays involving a $\chi_{c0}$ or $\chi_{c2}$ and
the corresponding ones with $J/\psi$ or $\chi_{c1}$ may thus shed
light on the relative importance of factorizable and nonfactorizable
contributions.

%------------------------------------------------------------------
\subsection{Penguins and decays into two light mesons}
\label{sec:light-light}
%------------------------------------------------------------------

Decays into two light mesons present some specifics due to the
presence of penguin operators. First, we have no meson candidates with
isospin $I=0$, which are a superposition of quark states $u\bar{u}$,
$d\bar{d}$, $s\bar{s}$. Penguin transitions lead to all three of them,
and one will always violate our condition that the spectator and the
emitted antiquark must have different flavors. Second, as remarked at
the end of Sect.~\ref{sec:suppress}, suppression mechanisms based on
the smallness of the decay constant $f_X$ are not effective if scalar
penguins occur in the factorization ansatz. Only spin suppression, and
the isospin suppression for the $b_1$, are still at work then.

It is nevertheless instructive to look at the relative importance of
the $S$ and $P$ operators in decays involving scalars or
pseudoscalars.  In the decays $\bar{B}^0\to a_0^+\, a_0^-$ or
$\bar{B}^0\to \pi^+(1300) \pi^-(1300)$ the scalar penguins come with
huge enhancement factors in the amplitudes
\begin{equation}
  r^{a_0} = \frac{2 m_{a_0}^2}{m_b\, (m_d-m_u)} , \hspace{2em}
  r^{\pi} = \frac{2 m_{\pi}^2}{m_b\, (m_d+m_u)} .
\label{r-def}
\end{equation}
Numerically, $r^{\pi(1300)}=85$, $r^{a_0(980)}=170$, and
$r^{a_0(1450)}=380$, to be compared with $r^\pi=1$ for the light pion,
where we have evolved the light quark masses (\ref{quark-masses}) up
to $\mu=m_b=4.4$~GeV. The strong scalar penguin can hence compete with
the current-current operators, even though its coefficient $a_6$
introduced in Sect.~\ref{sec:fac} is only of order of several
$10^{-2}$.

Using Eqs.~(\ref{eq:f_a0}), (\ref{eq:f_pi}), (\ref{r-def}), we can   
express the products $f_X r^X$ in terms of the (pseudo)scalar matrix
elements (\ref{pseudo-constants}) and (\ref{scalar-constants}), and 
observe that
\begin{equation}
f_{a_0} r^{a_0} \approx f_{\pi(1300)} r^{\pi(1300)} 
                \approx f_\pi r^\pi ,
\label{all-the-same}
\end{equation}
i.e., they are all roughly of the same size.  Since $\bar{B}^0\to
\pi^+\pi^-$ is driven by the color allowed tree level coefficient
$a_1$, naive factorization predicts the decay rate for $\bar{B}^0\to
a_0^+\, a_0^-$ to be small compared with the one for $\bar{B}^0\to
\pi^+\pi^-$. Namely, the ratio of their amplitudes is controlled by
the small parameters $f_{a_0} /f_\pi$ and
\begin{equation}
\frac{a_6}{a_1}\, \frac{f_{a_0} r^{a_0}}{f_\pi} =
\frac{a_6}{a_1}\,  r^\pi \,
  \frac{\vev{\bar{d} u}{a_0}}{
        \vev{\bar{d}\gamma_5\, u}{\pi}},
\label{peng-tree}
\end{equation}
corresponding to the tree level and scalar penguin contribution to
$\bar{B}^0\to a_0^+\, a_0^-$, respectively.  Analogous estimates can be
given for $\bar{B}^0\to \pi^+(1300)\, \pi^-(1300)$, and also for
$\bar{B}_s$ decays into $K^+ a_0^-$ and $K^+ \pi^-(1300)$ compared to
$K^+ \pi^-$.

The situation is different for decays where the emitted meson is a
kaon. The decay mode $\bar{B}^0\to \pi^+ K^-$ is penguin dominated
since its tree level contribution is CKM suppressed. With the analog
of (\ref{all-the-same}) for strange mesons one thus obtains similar
decay rates for $\bar{B}^0\to \pi^+ K^-$ and $\bar{B}^0\to \pi^+
K_0^{*-}$, where the latter receives most of its contribution from the
scalar penguin, as was pointed out in \cite{Chernyak:2001hs}. We
emphasize that, in contrast, naive factorization predicts
\mbox{${\cal{B}}(\bar{B}^0\to \pi^+ K^{*-}_2)=0$}.

%------------------------------------------------------------------

\subsection{Bottom baryon decays}

\begin{figure}
\begin{center}
\leavevmode
\epsfxsize=0.95\textwidth
\epsffile{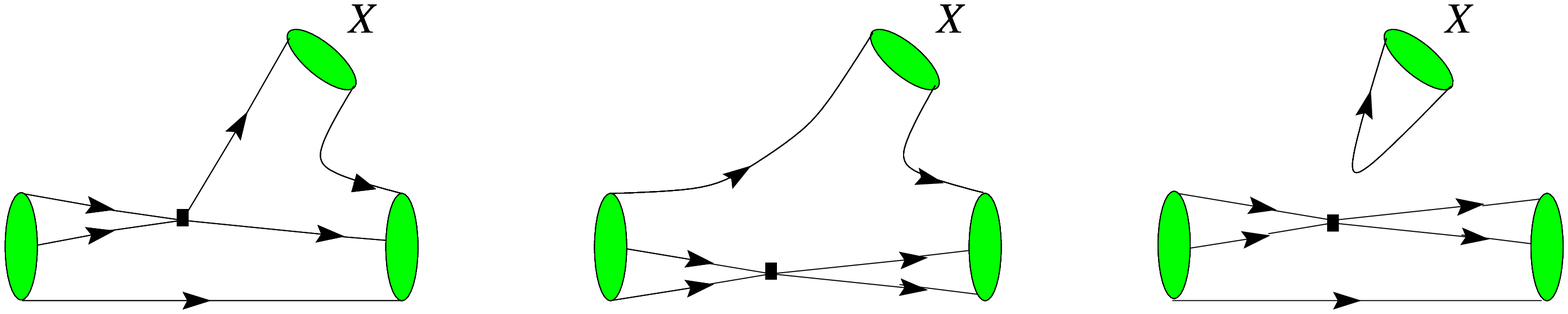}
\end{center}
\caption[]{The different topologies for pseudo-annihilation
contributions to $b$ baryon decays.}
\label{fig:bary}
\end{figure}

Bottom baryons provide a complementary field to study exclusive
hadronic decays, making more degrees of freedom such as polarization
accessible to experimental investigation. The most notable differences
between $q \bar{q}$ and $qqq$ bound states in the context of
factorization studies are the quark content and the role of
annihilation topologies.  Since the initial baryon can never be
completely annihilated by the operators in Eq.~(\ref{eq:heff}), we
call the corresponding topologies shown in Fig.~\ref{fig:bary}
pseudo-annihilation. For an overview of heavy baryon decays, we refer
to~\cite{Korner:1994nh}.

Let us adapt our ideas to study factorization and its breaking with
spin or decay constant suppression to the case of exclusive heavy
baryon decays.  Of course there is no background here from decays of
the $C\!P$ conjugated parent into the same final state, such as
discussed in Sect.~\ref{sec:flavor}. In order to ensure the formation
of the final state meson from the electroweak current, we must however
require that the spectator quarks in the baryon be different from the
quarks produced in the weak decay. In addition, we can only consider
bottom baryons that have weak decays and do not dominantly decay
strongly or electromagnetically. A possible decay channel is for
instance $\Lambda_b \to \Lambda_c D_{sJ}^-$. Also, Mannel et
al.~\cite{Mannel:1991vg} have mentioned that the $\Omega_b$ might only
have electroweak decays. In Table~\ref{tab:baryons} we list the decays
of these two baryons for which it is assured that the meson cannot
pick up a spectator, so that corrections to factorization can be
studied with our method.

\begin{table}[tb]
\begin{center}
\renewcommand{\arraystretch}{1.2}
\begin{tabular}{lccccccc} \hline \hline
example decays & \multicolumn{3}{c}{factorizing contribution} &
        \multicolumn{4}{c}{pseudo-annihilation} \\
 & quark level & tree & peng. & ~ & tree & peng. & \\ \hline
$\Lambda_b\to \phantom{\Lambda_c K_0^{*-},\;}
  n D_2^{*0}$ &
  $udb\to ud(c\bar{u}d)$ & 
  $\lambda^2$ & & & $\lambda^2$\\
$\phantom{\Lambda_b}\to \Lambda_c K_0^{*-},\, \Lambda D_2^{*0}$ & 
  $\phantom{udb}\to ud(c\bar{u}s)$ & 
  $\lambda^3$ & & & $\lambda^3$ \\
$\phantom{\Lambda_b}\to p\, D_{sJ}^-$ &
  $\phantom{udb}\to ud(u\bar{c}s)$ & 
  $\lambda^3$ & \\
$\phantom{\Lambda_b}\to \Lambda_c D_{sJ}^-,\, \Lambda \chi_{c0}$ &
  $\phantom{udb}\to ud(c\bar{c}s)$ & 
  $\lambda^2$ & $\lambda^2$ & &
  $\lambda^4$ & $\lambda^2$  \\
$\phantom{\Lambda_b}\to \phantom{\Lambda_c D_{sJ}^-,\;} 
  n\, \chi_{c0}$ &
  $\phantom{udb}\to ud(c\bar{c}d)$ & 
  $\lambda^3$ & $\lambda^3$ & &
  $\lambda^3$ & $\lambda^3$ \\
\hline
$\Omega_b\to \Omega_c a_0^{-},\, \Xi^- D_2^{*0}$ &
  $ssb\to ss(c\bar{u}d)$ &
  $\lambda^2$ & \\
$\phantom{\Omega_b}\to \phantom{\Omega_c K_0^{*-},\;}
  \Omega\, D_2^{*0}$ &
  $\phantom{ssb}\to ss(c\bar{u}s)$ &
  $\lambda^3$ & \\
$\phantom{\Omega_b}\to \phantom{\Omega_c K_0^{*-},\;}
  \Omega\,  D_2^{*0}$ & 
  $\phantom{ssb}\to ss(u\bar{c}s)$ &
  $\lambda^3$ &\\
$\phantom{\Omega_b}\to \Xi^0 D_2^{*-},\, \Xi^- D_2^{*0}$ &
  $\phantom{ssb}\to ss(u\bar{c}d)$ &
  $\lambda^4$ & \\
$\phantom{\Omega_b}\to \phantom{\Omega_c K_0^{*-},\;}
  \Omega\,  \chi_{c0}$ & 
  $\phantom{ssb}\to ss(c\bar{c}s)$ &
  $\lambda^2$ & $\lambda^2$ & & 
    & $\lambda^2$  \\
$\phantom{\Omega_b}\to \Omega_c D_2^{*-},\, \Xi^- \chi_{c0}$ &
  $\phantom{ssb}\to ss(c\bar{c}d)$ &
  $\lambda^3$ & $\lambda^3$ & & 
    & $\lambda^3$  \\
$\phantom{\Omega_b}\to \phantom{\Omega_c K_0^{*-},\;}
  \Omega\,  a_2^0$ &
  $\phantom{ssb}\to ss(u\bar{u}s)$ &
  $\lambda^4$ & $\lambda^2$ & & 
   &   \\
$\phantom{\Omega_b}\to \phantom{\Omega_c K_0^{*-},\;}
  \Xi^- a_2^0$ &
  $\phantom{ssb}\to ss(u\bar{u}d)$ &
  $\lambda^3$ & $\lambda^3$ & &
    & $\lambda^3$  \\
\hline \hline
\end{tabular}
\end{center}
\caption{\label{tab:baryons} Flavor structure of the decay modes of
the $\Lambda_b$ and the $\Omega_b$ for which the emitted meson must
originate from the weak current. The mesons can be replaced by others
from Table~\protect\ref{tab:friends} with the same flavor
structure. The second column gives the quark composition of the final
state, with the three quarks originating from the electroweak vertex
enclosed in brackets. The third and fourth columns give the power of
the Wolfenstein parameter $\lambda$ in the decay amplitude for $W$
exchange and penguin operators, respectively, and the corresponding
information for pseudo-annihilation contributions is given in the last
two columns.}
\end{table}

%%%%%%%%%%%%%%%%%%%%%%%%%%%%%%%%%%%%%%%%%%%%%%%%%%%%%%%%%%%%%%%%%%%
\section{Escaping suppression by factorization breaking}
\label{sec:break}
%%%%%%%%%%%%%%%%%%%%%%%%%%%%%%%%%%%%%%%%%%%%%%%%%%%%%%%%%%%%%%%%%%%

The idea of this paper is to study the size and pattern of corrections
to factorization in an environment where they are not ``hidden''
behind a larger factorizing piece. Without giving an exhaustive
discussion of nonfactorizing contributions, we now consider two of
them and see why the various suppression mechanisms discussed in
Sect.~\ref{sec:suppress} do not apply.

%------------------------------------------------------------------
\subsection{Nonfactorizing gluon exchange}
\label{sec:non-fact}
%------------------------------------------------------------------

\begin{figure}
\begin{center}
\leavevmode
\epsfxsize=0.9\textwidth
\epsffile{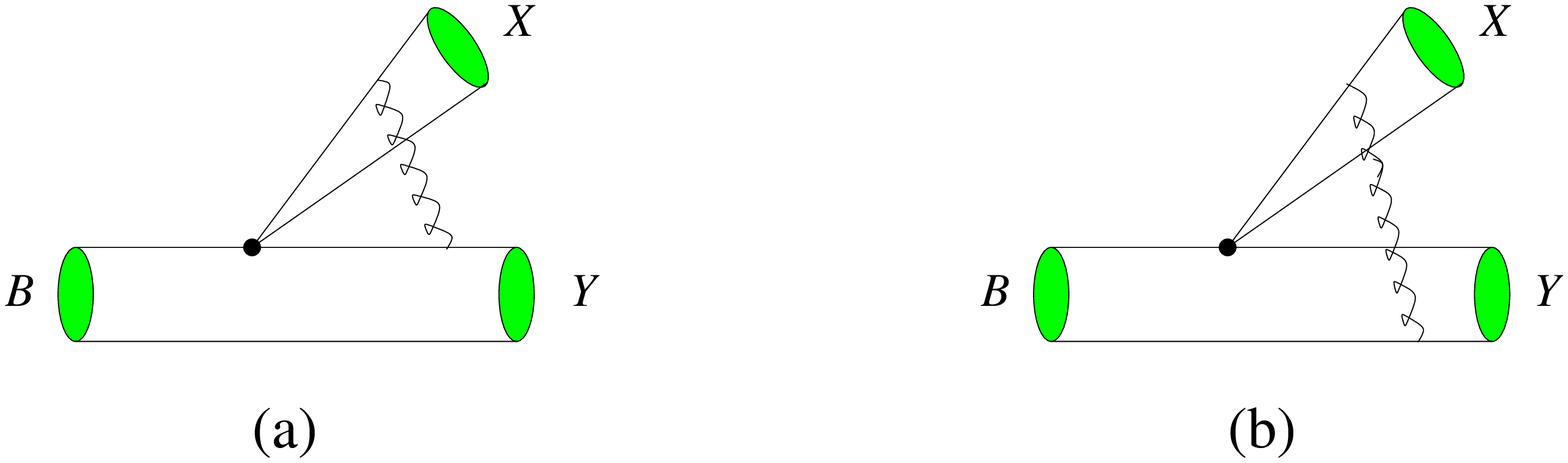}
\end{center}
\caption[]{Examples of diagrams with gluon exchange breaking naive
factorization: (a) vertex correction, (b) spectator interaction.}
\label{fig:break}
\end{figure}

Naive factorization is broken by strong interactions of the quarks
originating from the $b$ decay. In the language of quarks and gluons
they correspond to diagrams like the ones in
Fig.~\ref{fig:break}. What is important in our context is that such
contributions no longer involve the matrix elements of the meson $X$
with the local currents $V$, $A$, $S$, $P$. As a consequence the
suppression mechanisms based on the smallness of the decay constant
$f_X$ are not effective. Also, one or several gluons absorbed by the
quark-antiquark pair that will form $X$ can transfer both helicity and
orbital angular momentum, so that the spin of $X$ is no longer
restricted to be 0 or 1.

Note that these arguments are independent on whether the internal
lines in the diagrams of Fig.~\ref{fig:break} have large virtualities
or not. In the first case the corresponding contributions can be
calculated in perturbation theory. In Sect.~\ref{sec:bbns} we will
analyze them in the QCD factorization approach and explicitly see that
our suppression mechanisms are no longer operative.

If the internal lines in these diagrams are not hard, perturbation
theory is not reliable, and other descriptions of the corresponding
reactions might be more adequate. If one treats them for instance as
hadronic rescattering, there is again no reason why final state mesons
$X$ with small decay constants or higher spin should be suppressed.

%------------------------------------------------------------------
\subsection{Annihilation contributions} 
%------------------------------------------------------------------

Annihilation diagrams are another important contribution violating
naive factorization. In Tables~\ref{tab:flavor} and
\ref{tab:flavor-two} we have indicated the channels where they can
occur, either from tree level $W$ exchange or from penguin
operators. An indication of the importance of annihilation could be
obtained from data by comparing decays with and without such
contributions that are otherwise similar, for instance $\bar{B}^0\to
D^+ a_0^-$ and $\bar{B}_s\to D_s^+ a_0^-$, or $B^-\to \pi^0 D_{sJ}^-$
and $B^-\to \eta D_{sJ}^-$.

\begin{figure}
\begin{center}
\leavevmode
\epsfxsize=0.95\textwidth
\epsffile{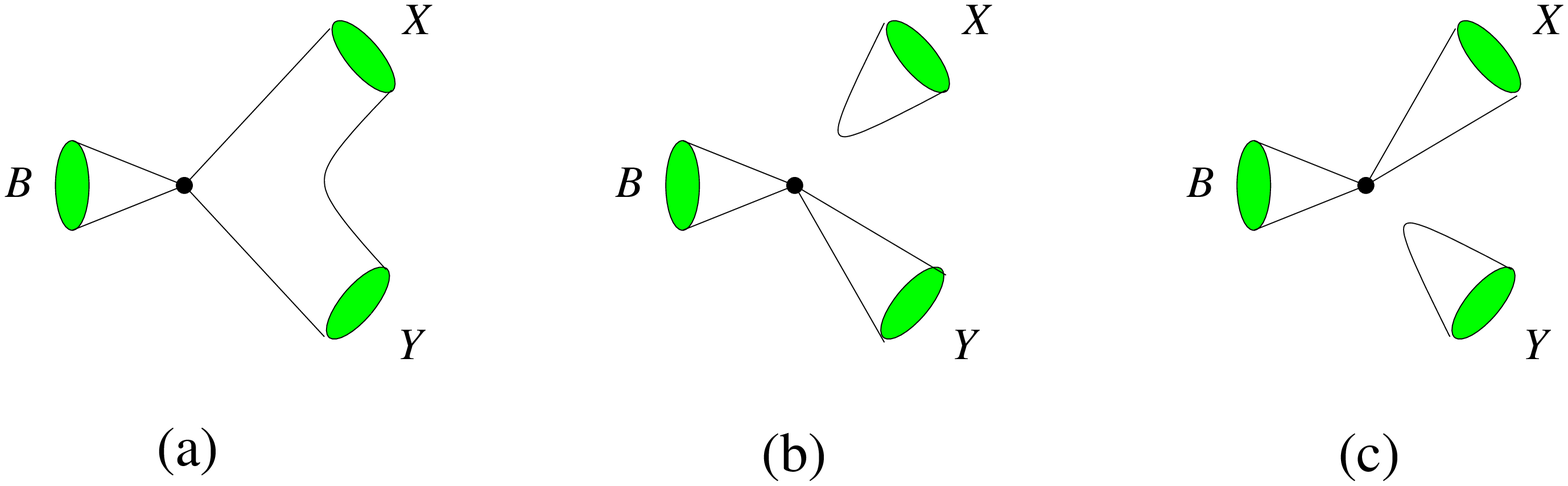}
\end{center}
\caption[]{The different topologies for annihilation contributions to
decays $B\to Y X$, where $X$ is one of our candidate mesons given
in Table~\ref{tab:friends}.}
\label{fig:anni}
\end{figure}

Depending on the flavor structure of the decay there are three
annihilation topologies. The one shown in Fig.~\ref{fig:anni}a is for
example relevant for $\bar{B}^0\to D^+ a_0^-$ and for $B^-\to \pi^-
\bar{K}_2^{*0}$. In this case only one of the quarks forming our
candidate meson $X$ originates from the effective weak vertex, so that
our suppression mechanisms are not relevant. Notice that this holds
true irrespective of whether the interactions between the
$q\bar{q}$-pair and the quarks attached to the decay vertex are under
perturbative control or not, a point that is controversial in the
literature \cite{Beneke:2001ev,Keum:2001wi}.

The two remaining topologies correspond to Zweig forbidden
contributions. Neglecting electromagnetic interactions, they require
that the meson formed from the $q\bar{q}$-pair has isospin $I=0$. In
Fig.~\ref{fig:anni}b, relevant for our decays into charmonium, neither
of the quarks from the annihilation vertex enters in $X$, so that our
suppression mechanisms again do not apply. Finally, there is the case
of Fig.~\ref{fig:anni}c, where $X$ is formed from the quarks of the
decay vertex. Our suppression mechanisms only apply here if the
$\bar{q}q$-pair interacts solely with the quarks in the $B$ meson, but
not with the ones forming the meson $X$.

The topologies for pseudo-annihilation contributions in $b$ baryon
decays have already been shown in Fig.~\ref{fig:bary}. All of them
circumvent suppression.

%%%%%%%%%%%%%%%%%%%%%%%%%%%%%%%%%%%%%%%%%%%%%%%%%%%%%%%%%%%%%%%%%%%
\section{The case of QCD factorization}
\label{sec:bbns}
%%%%%%%%%%%%%%%%%%%%%%%%%%%%%%%%%%%%%%%%%%%%%%%%%%%%%%%%%%%%%%%%%%%

In the QCD factorization approach developed by Beneke et
al.~\cite{Beneke:1999br,Beneke:2000ry,Beneke:2001ev}, those
corrections to naive factorization that are dominated by hard gluon
exchange are calculated in perturbation theory, all other
contributions are found to be power suppressed in $1/m_b$. The
physical mechanism underlying these results is color transparency of
the meson ejected by the effective weak current. QCD
factorization can therefore not be applied to decays where a $D$ meson
with its highly asymmetric quark-antiquark configurations is emitted
from the current, and we will thus not consider the corresponding
channels in this section.

As QCD factorization relies on color transparency, it also requires
the emitted meson to be fast in the $B$ rest frame. Certain types of
power corrections will therefore increase with the mass $m_X$ of the
emitted meson. One can reduce the bias due to such effects by
comparing our suppressed decays with unsuppressed channels involving
mesons of similar mass. Examples are the $\rho(770)$, $a_1(1260)$, or
$\rho(1450)$ for isospin-one mesons, and the $K^*$ and $K_1$
resonances for the strange sector.

%------------------------------------------------------------------

\subsection{Distribution amplitudes}
\label{sec:da}

The most important effect of radiative corrections in our context is
that the currents like $\vev{\bar{d}(0) \gamma^\mu (1-\gamma_5)
u(0)}{X}$ occurring in naive factorization become nonlocal. The
leading configurations in $1/m_b$ involve light-like separations $z$
and are parameterized by meson distribution amplitudes. In light-cone
gauge, they read for a $d\bar{u}$ meson
\begin{equation}
\vev{\bar{d}(z)\, \gamma^{\mu}(\gamma_5)
                  u(-z)}{X(q,J_3=0)} \Big|_{z^2=0} =
-i q^\mu \int_0^1 du\, e^{i(2u-1)\, q\cdot z}\, \varphi(u) + \ldots
\label{da-def}
\end{equation}
to twist-two accuracy, where the $\ldots$ stand for terms of twist
three and higher, which contribute to hard processes only at the power
correction level. The Dirac matrices $\gamma^\mu$ and $\gamma^{\mu}
\gamma_5$ are to be taken for mesons with natural and unnatural
parity, $P=(-1)^J$ and $P=(-1)^{J+1}$, respectively. The variable $u$
gives the momentum fraction carried by the quark in the meson $X$, a
natural frame of reference in our case being the $B$ rest
frame. Notice that the twist-two distribution amplitudes involving the
vector and axial currents select the polarization state with zero
angular momentum $J_3$ along $\vec{q}$ in that frame. We remark that
our subsequent discussion remains valid if instead of a meson $X$ one
considers a continuum final state with appropriate quantum numbers as
discussed in Sect.~\ref{sec:continuum}. In this case $\varphi$ is to
be replaced with a generalized distribution amplitude
\cite{Muller:1994fv}, defined as in Eq.~(\ref{da-def}) with the
appropriate replacement of state vectors $|X\rangle$.

One easily sees that the lowest moment of $\varphi$ in $u$ gives back
the local currents, so that for mesons with spin 0 or 1 one recovers
the decay constant,
\begin{equation}
  \int_0^1 du\, \varphi(u) = f_X .
\label{phi-norm}
\end{equation}
The nonlocal currents in Eq.~(\ref{da-def}) have however matrix
elements for mesons of any spin $J$. Taylor expanding the bilocal
operators around $z=0$ gives in fact operators with arbitrarily high
numbers of partial derivatives between $\bar{d}(0)$ and $u(0)$, and
thus with arbitrarily high spin. In other words the lowest moment
$\int du\, \varphi(u)$ of the distribution amplitude, projecting out
the local $V$ or $A$ current, vanishes for mesons with $J\ge 2$, but
not the function $\varphi(u)$ itself.  Hence the production of mesons
with spin 2 and higher is no longer forbidden at the level of
$\alpha_s$ corrections to naive factorization. We thus find an
explicit realization of our arguments in Sect.~\ref{sec:non-fact}:
gluon exchange such as in Fig.~\ref{fig:break} indeed makes the
production of higher-spin mesons possible.

Let us now see what becomes of the other suppression mechanisms we
discussed in Sect.~\ref{sec:suppress}. Charge conjugation invariance
implies that the distribution amplitudes $\varphi_{X_0}$ for the
\emph{neutral} $a_0$ and $b_1$ mesons are odd under the exchange of
quark and antiquark momenta, $\varphi_{X^0}(u) = -
\varphi_{X^0}(1-u)$, so that their first moment ~(\ref{phi-norm}) is
zero. In the exact isospin limit, the distribution amplitudes for the
charged and neutral mesons in an isotriplet are the same, so that
with Eq.~(\ref{phi-norm}) the decay constants of the charged $a_0$ and
$b_1$ have to vanish, as already seen in Sect.~\ref{sec:suppress}. In
the real world, the part of the distribution amplitude that is even
under $u\to 1-u$ is therefore small for the charged $a_0$ and $b_1$,
i.e., $\varphi_{X^-}(u) + \varphi_{X^-}(1-u) \sim m_d-m_u$. This does
however not restrict the odd part of $\varphi_{X^-}(u)$, which can be
comparable in size to the distribution amplitudes of, say, the $\pi$
or the $\rho$. Considering the distribution amplitude of the $K_0^*$
and using SU(3) symmetry we see that its even part is suppressed by
$m_s-m_u$, but not its odd part.

Along the same lines of reasoning, we find that up to isospin breaking
effects the distribution amplitudes for the charged heavy pions are
even under $u\to 1-u$. The lowest moment $\int du\,
\varphi_{\pi(1300)}(u)$ is small of order $m_u+m_d$ according to
Eqs.~(\ref{eq:f_pi}) and (\ref{phi-norm}), but there is no such
restriction on higher even moments such as $\int du\, (2u-1)^2\,
\varphi_{\pi(1300)}(u)$.  We thus see that neither spin nor any of our
other suppression mechanisms applies at the level of $\alpha_s$
corrections.

It is useful to expand the distribution amplitude in Gegenbauer
polynomials $C_n^{3/2}$, which are the eigenfunctions of the
leading-order evolution equation \cite{Lepage:1979zb} for quark
distribution amplitudes. In order to achieve a uniform notation for
mesons with different spins we write
\begin{equation}
\varphi(u;\mu) = f^\varphi\, 6 u(1-u) \Big[ B_0 + \sum_{n=1}^{\infty}
              B_n(\mu)\, C_n^{3/2}(2u-1) \Big] ,
\label{gegenbau}
\end{equation}
where we have explicitly displayed the dependence on the factorization
scale $\mu$. For mesons $X$ with spin 0 or 1, we have $B_0=1$ and
$f^\varphi$ is just the decay constant $f_X$, whereas for mesons with
$J\ge 2$ we have $B_0=0$. We will in the following only use products
$f^\varphi B_n$ so that we need not specify the separate
normalizations of $f^\varphi$ and $B_n$ in that case. {}From our above
discussion it follows that for our candidate mesons one or more of the
lowest coefficients in the expansion (\ref{gegenbau}) are either zero
or small of order $m_d-m_u$, $m_d+m_u$, or $m_s-m_u$. In the Appendix
we will estimate the orders of magnitude of the leading coefficients
to be
\begin{eqnarray}
|f^\varphi B_1|_{a_0, b_1, a_2, K_0^*, K_2^*} 
      &\approx& 75~\mbox{MeV} ,
\nonumber \\
|f^\varphi B_2|_{\pi(1300), \pi_2, \rho_3} &\approx& 50~\mbox{MeV} ,
\label{eq:fB12}
\end{eqnarray}
evaluated at the renormalization scale $\mu=m_b=4.4$~GeV.  

%------------------------------------------------------------------

\subsection{Decays into a $D$ and a light meson $X$}
\label{sec:heavy-light-qcd}

For these channels the only diagrams one needs to consider at leading
order in $1/m_b$ are vertex corrections such as in
Fig.~\ref{fig:break}a. The result of the $O(\alpha_s)$ calculation for
the matrix element of the effective weak Hamiltonian can be written as
the sum of
\begin{eqnarray}
\langle D^+ X^-| {\cal H}_{\mathrm{eff}} | 
      \bar{B}^0\rangle^{\mathrm{corr}} &=&
-i\, \frac{G_F}{\sqrt{2}} V_{cb}^{\phantom{*}} V_{ud}^* \;
   a_1^{\mathrm{corr}} f^\varphi\, q_\mu
   \langle D^+ |\bar{c} \gamma^\mu b |\bar{B}^0\rangle ,
\nonumber \\
\langle D^{*+} X^-| {\cal H}_{\mathrm{eff}} | 
      \bar{B}^0\rangle^{\mathrm{corr}} &=&
 i\, \frac{G_F}{\sqrt{2}} V_{cb}^{\phantom{*}} V_{ud}^* \;
   a_1^{\mathrm{corr}} f^\varphi\, q_\mu
   \langle D^{*+} |\bar{c}\gamma^\mu\gamma_5 b
       |\bar{B}^0\rangle ,
\label{one-loop}
\end{eqnarray}
and the contribution of naive factorization, taken with a coefficient
$a_1^{\mathrm{fact}}$ which equals the coefficient $a_1$ of
Sect.~\ref{sec:fac} evaluated at next-to leading order, up to a small
term removing the renormalization scheme dependence. For details we
refer to Eqs.~(95) and (96) of \cite{Beneke:2000ry}. The coefficients
$a_1^{\mathrm{corr}}$ are given by
\begin{equation}
f^\varphi a_1^{\mathrm{corr}} = \frac{\alpha_s(\mu)}{4\pi}\,C_2(\mu)\,
  \frac{C_F}{N_c}\, \int_0^1 du\, F(u, \pm z)\, 
  \varphi(u;\mu) ,
\end{equation}
where the function $F(u,z)$ with $z=m_c/m_b$ can be found
in~\cite{Beneke:2000ry}. Its second argument is $+z$ for decays into
$D$ and $-z$ for decays into $D^*$, with $z=m_c/m_b$, so that 
$a_1^{\mathrm{corr}}$ depends on both mesons in the final state. Its
dependence on the distribution amplitude of $X$ can be expressed as
\begin{equation}
a_1^{\mathrm{corr}}(\mu) = a_1^{(0)}(\mu)\, B_0 + 
      \sum_{n=1}^\infty a_1^{(n)}(\mu)\, B_n(\mu) .
\label{expand-a1}
\end{equation}
The first few coefficients $a_1^{(n)}$ are listed in
Table~\ref{tab:gegenbau}. We see that they are small compared with 1,
and that they tend to decrease with $n$. They depend substantially on
the renormalization scale $\mu$. One finds
\begin{equation}
\frac{a_1^{(n)}(m_b/2)}{a_1^{(n)}(m_b)} \approx 
\frac{a_1^{(n)}(m_b)}{a_1^{(n)}(2m_b)} \approx 2,
\label{variation}
\end{equation}
a falloff mostly due to the Wilson coefficient $C_2$. The Gegenbauer
coefficients $B_n$ also decrease with the factorization scale,
although by less than a factor 1.2 for $B_1$ and $B_2$ when $\mu$ is
varied between $m_b/2$ and $m_b$ or between $m_b$ and $2m_b$. The
effects of this dependence on $\mu$ are quite mild for decays where
most of the result is due to the Born level term $f^\varphi
a_1^{\mathrm{fact}}$, but not for our decays where this contribution
is absent or suppressed by a small value of $f^\varphi$. A more stable
prediction would require the inclusion of $O(\alpha_s^2)$ corrections.
Since they involve again the color allowed Wilson coefficient $C_1$
they may actually not be small compared with the $O(\alpha_s)$ terms.

\begin{table}
\begin{center}
\renewcommand{\arraystretch}{1.2}
\begin{tabular}{lccccc} \hline \hline
 & $a_1^{(0)}$ & $a_1^{(1)}$ & $a_1^{(2)}$
 & $a_1^{(3)}$ & $a_1^{(4)}$  \\ \hline
$\bar{B}^0\to D^+ X^-$ & $12.3+12.7 i$ & $-8.1+17.8 i$ & 
  $1.7-0.8 i$ & $-2.3+4.9 i$ & $\hfill 0.3- 0.5 i$ \\
$\bar{B}^0\to D^{*+} X^-$ & $11.7+8.4 i \hfill$ & $-9.5+16.0 i$ & 
  $0.3-1.2 i$ & $-3.2+5.0 i$ & $-0.2-0.2 i$ \\
\hline \hline
\end{tabular}
\end{center}
\caption{\label{tab:gegenbau} The first few coefficients $a_1^{(n)}$
of the expansion (\protect\ref{expand-a1}) in units of $10^{-3}$. They
are evaluated at renormalization scale $\mu=m_b=4.4$~GeV with input
parameters $\Lambda^{(5)}_{\overline{\mathrm{MS}}}=220$ MeV and
$m_c/m_b=0.3$. With the same parameters one has
$a_1^{\mathrm{fact}}=1.039$.}
\end{table}

Whereas $a_1^{\mathrm{fact}}$ is real valued, we observe from
Eq.~(\ref{expand-a1}) and Table~\ref{tab:gegenbau} that
$a_1^{\mathrm{corr}}$ is complex. The strong phases of unsuppressed
channels like $\bar{B}^0\to D^+ \pi^-$ are thus small in QCD
factorization. On the contrary, they can be sizeable in our suppressed
decays, where $\alpha_s$ contributions are essential.

%------------------------------------------------------------------

\subsection{Decays into charmonium}

The radiative corrections to factorization for decays into charmonium
states have not been calculated yet. The following observation
\cite{Beneke:2000ry} is however relevant in our context. The naive
factorization formula for these decays involves the color suppressed
coefficient $a_2$ and color suppressed penguins, but at the level of
loop corrections the color allowed coefficient $C_1$ will come
in. Hence the $O(\alpha_s)$ terms will probably be sizeable compared
with the naive factorization result. This expectation is supported by
an analysis of inclusive $B$ decays into charmonium
\cite{Beneke:1999ks}.  Whereas naive factorization forbids the decays
of Table~\ref{tab:flavor-two} into a $\chi_{c0}$ or a $\chi_{c2}$
instead, one may then expect that within QCD factorization their
branching ratios are not much smaller or even of similar size than for
the corresponding decays into $J/\Psi$ or $\chi_{c1}$.

%------------------------------------------------------------------

\subsection{Decays into two light mesons}

For these channels, not only vertex corrections need to be considered,
but also so-called penguin contractions (cf.\ Fig.~7 of
\cite{Beneke:2000ry}) and hard interactions with the spectator quark
from the $B$ as shown in Fig.~\ref{fig:break}b. The latter involve the
twist-two distribution amplitudes $\varphi_Y$ and $\varphi_X$ of both
final state mesons. Mesons with $J\ge 1$ have a second twist-two
distribution amplitude~\cite{Ball:1998fj} involving the nonlocal
tensor current $\bar{d}(z)\, \sigma^{\mu\nu} u(-z)$, which describes
states with helicity \mbox{$J_3=\pm 1$} and can also contribute now.

Beyond the level of leading twist contributions, several power
corrections have been considered in the literature
\cite{Beneke:2001ev,Du:2001ns}. A particular type of correction that
is numerically not suppressed occurs for final-state mesons with spin
$J=0$ and involves their twist-three distribution amplitudes defined
from the nonlocal (pseudo)scalar and tensor currents
\cite{Braun:1990iv}. Their contribution to the amplitude relative to
the twist-two radiative corrections is controlled by the ratio $r$
defined as in (\ref{r-def}), which is formally of order $1/m_b$ but
numerically not small. With the quark mass values we use, one has
$r_\pi = r_K =1$, and the corresponding twist-three radiative
corrections have size similar to the leading-twist ones. This is also
the case for the $a_0$, $\pi(1300)$, and $K_0^*$ mesons we are
considering here.  Indeed, the twist-two distribution amplitudes are
comparable in size for $a_0$, $\pi(1300)$, $K_0^*$, and $\pi$, $K$, as
follows from comparing our estimates (\ref{eq:fB12}) with $f_\pi$ or
$f_K$.  The same holds for the respective twist-three distribution
amplitudes, which are controlled by the local (pseudo)scalar matrix
elements in Eqs.~(\ref{pseudo-constants}) and
(\ref{scalar-constants}). We recall that the twist-three pieces just
discussed can only be \emph{estimated} in QCD factorization because
they lead to logarithmic divergences at the endpoints of the
distribution amplitudes.

Logarithmic endpoint divergences also appear in annihilation
contributions, which in the power counting scheme of QCD factorization
are again $1/m_b$ corrections, even for those terms involving only
twist-two distribution amplitudes. In a recent study of decays $B\to
\pi\pi$ and $B\to \pi K$ Beneke et al.~\cite{Beneke:2001ev} estimated
annihilation contributions to be moderate corrections to the leading
terms computed in the QCD factorization framework, giving a benchmark
number of 25\% in the branching ratio, although with large
uncertainties. Notice that the small parameter controlling the
relative weight of annihilation and leading contributions in their
calculation is
\begin{equation}
\frac{f_B f_Y}{(m_B^2-m_Y^2) F_0^{B\to Y}(m_X^2)} ,
\end{equation}
which depends only mildly on the mass of the emitted meson through the
$B\to Y$ transition form factor. We take this as an indication that 
the importance of annihilation contributions is not primarily driven 
by the size of $m_X$. 

We expect then that in decays such as $\bar{B}^0\to a_0^+ a_0^-$ the
hard nonfactorizable terms calculable in QCD factorization, and
possibly also the power corrections estimated there should be small
compared to the amplitude for $\bar{B}^0\to \pi^+\pi^-$, which is
dominated by the contribution of the large coefficient $a_1$.

For penguin dominated decays like $B\to \pi K$, the overall size of
corrections found in Ref.~\cite{Beneke:2001ev} is not so small
compared with the result of the naive factorization formula. With the
``designer'' modes $\bar{B}^0 \to \pi^+ K_2^{*-}$ and $B^-\to \pi^-
\bar{K}_2^{*0}$ we can isolate such nonfactorizable terms. Among
these, annihilation contributions from scalar penguin operators are of
special interest. In QCD factorization they are a power correction,
but not in the PQCD approach, where Keum et al.~\cite{Keum:2001wi}
found that they contribute at order one and with a large phase to the
$B\to \pi K$ decay amplitudes. Comparing the branching ratios of the
above decays into $K_2^{*}$ with those into a $K$ or $K^*$ can show
whether scalar penguin annihilation contributions are indeed large.

%%%%%%%%%%%%%%%%%%%%%%%%%%%%%%%%%%%%%%%%%%%%%%%%%%%%%%%%%%%%%%%%%%%
\section{Branching ratio estimates for decays $B\to D X$}
\label{sec:branching}
%%%%%%%%%%%%%%%%%%%%%%%%%%%%%%%%%%%%%%%%%%%%%%%%%%%%%%%%%%%%%%%%%%%

We present now our numerical estimates of the branching ratios for
decays $\bar{B}^0 \to D^+ X^-$, where $X$ is one of $a_0$, $a_2$,
$b_1$, $\pi(1300)$, $\pi_2$, $\rho_3$, or $K_0^*(1430)$, $K_2^*$. As
discussed in Sect.~\ref{sec:heavy-light-qcd}, these channels receive
hard gluon corrections to naive factorization, which can be calculated
in the QCD factorization approach.

We recall that for the $a_2$, $\pi_2$, $ \rho_3$, and $K_2^*$, the
tree term proportional to $a_1^{\mathrm{fact}}$ is absent. The
contribution from the $\alpha_s$ correction $a_1^{\mathrm{corr}}$
given in (\ref{one-loop}) involves a contraction of $q_\mu$ with the
matrix element parameterized as
\begin{eqnarray}\label{eq:ff1}
\langle D(p^\prime) | \bar c \gamma^\mu b | \bar{B}(p)\rangle  
&=&  F_1(q^2) \left\{
(p+p^\prime)^\mu - \frac{m_B^2-m_D^2}{q^2}\,  q^\mu \right\} +
F_0(q^2)\, \frac{m_B^2-m_D^2}{q^2}\, q^\mu , 
\end{eqnarray}
where $q=p-p^\prime$. One easily sees that the $\alpha_s$
contributions always pick up the form factor $F_0$, independent of the
spin of $X$.
Thus, the decay rate for our candidates $X$, as well as for any other
spin zero meson like the $\pi$, can be written as
\begin{eqnarray}
\Gamma(\bar{B}^0 \to D^+ X^-)&=&
\frac{G_F^2}{16 \pi}\, \frac{(m_B^2-m_D^2)^2}{m_B^2}\, p_X \,
 \Big|V^{\phantom{*}}_{cb} V_{ud}^*\Big|^2\; \Big|\, f^\varphi 
   (a_1^{\mathrm{fact}} + a_1^{\mathrm{corr}})\,  F_0(m_X^2) \,\Big|^2 ,
\end{eqnarray}
where $p_X$ denotes the magnitude of the three-momentum of $X$
in the $B$ rest frame.

We normalize the rate of the decays into $X$ to the unsuppressed one
into a light pion. This has the advantage that the CKM factors cancel
(except for the strange mesons) and that we can use
the measured branching ratio for  $\bar{B}^0 \to D^+ \pi^-$ decays,
where naive factorization works well
\cite{Neubert:1997uc,Beneke:2000ry}.
For the ratio of decay rates we have the simple expression
\begin{equation}
\frac{\Gamma(\bar{B}^0 \to D^+ X^-)}{\Gamma(\bar{B}^0 \to D^+ \pi^-)} 
\approx
\left|\,
   \frac{f^\varphi (a_1^{\mathrm{fact}} + a_1^{\mathrm{corr}})}{
        f_\pi a_1^{\mathrm{fact}}} \,\right|^2  ,
\label{eq:rate}
\end{equation}
where for simplicity we have neglected the term $a_1^{\mathrm{corr}}$
for the $\pi^-$, where its effect is of order 1\%~to~2\%.
Eq.~(\ref{eq:rate}) has corrections due to phase space and the
evaluation of the form factor $F_0$ at different momentum transfer
$q^2$, which go in opposite directions. The relevant $q^2$ ranges from
roughly 1~GeV$^2$ for the $a_0(980)$ to 3~GeV$^2$ for the $\pi_2$ and
$\rho_3$, to be compared with $q^2\approx 0$ for the $\pi$.  We have
checked that the relation (\ref{eq:rate}) is affected by these mass
effects by not more than 10\% to 15\%. This is sufficient in our
context, given the dominant theoretical uncertainty of our calculation
hidden in the decay constants and distribution amplitudes.  For
$X=K_0^*(1430)$, $K_2^*$ we have to include a CKM factor
$|V_{us}/V_{ud}|^2$, which is known to a good precision. In our
numerical analysis we take $|V_{us}/V_{ud}|=0.23$ from
\cite{Groom:2000in}.

We evaluate (\ref{eq:rate}) using the expansion (\ref{expand-a1}) with
the coefficients $a_1^{(n)}$ of Gegenbauer moments in Table
\ref{tab:gegenbau}, the estimates (\ref{eq:fB12}) for the Gegenbauer
moments $B_n$, and the maximal values of the decay constants in
(\ref{malt}) to (\ref{light-pi}).  The $b_1$ can have a small decay
constant, for which we are not aware of any information in the
literature, and in our calculation we set it to zero.
Before presenting the branching ratios let us study the magnitude of
the individual terms entering in (\ref{eq:rate}). We have $f_\pi
a_1^{\mathrm{fact}} = 136$~MeV for the pion, and for the mesons
with leading Gegenbauer moments $f^\varphi B_1$ and $f^\varphi B_2$,
we respectively obtain
\begin{eqnarray}
|f^\varphi a_1^{\mathrm{corr}} \,|_{a_0,b_1,a_2,K_0^*,K_2^*} 
  &=& 1.5~\mbox{MeV} , \nonumber \\
|f^\varphi a_1^{\mathrm{corr}} \,|_{\pi(1300),\pi_2,\rho_3}
  &=& 0.1~\mbox{MeV} ,
\end{eqnarray}
at renormalization scale $\mu=m_b$. The second term is tiny mainly
because of the small coefficient $a_1^{(2)}$ from the one-loop
calculation. Since our estimate in the Appendix does not yield the
sign of $f^\varphi B_n$, we have a twofold ambiguity when adding
$f^\varphi a_1^{\mathrm{corr}}$ to $f^\varphi a_1^{\mathrm{fact}}$,
and find
\begin{eqnarray}
f_{a0(980)}\, a_1^{\mathrm{fact}} &=& 2.6 \mbox{~MeV} , \hspace{3em}
|f^\varphi (a_1^{\mathrm{fact}}+a_1^{\mathrm{corr}}) \,|_{a0(980)}
  \hspace{0.55em} \,=\;
              2.4 \mbox{~or~} 3.5 \mbox{~MeV} , \nonumber \\
f_{a0(1450)}\, a_1^{\mathrm{fact}} &=& 0.7 \mbox{~MeV} , \hspace{3em}
|f^\varphi (a_1^{\mathrm{fact}}+a_1^{\mathrm{corr}}) \,|_{a0(1450)}
  \hspace{0.25em} \,=\; 
              1.3 \mbox{~or~} 1.9 \mbox{~MeV} , \nonumber\\
f_{\pi(1300)}\, a_1^{\mathrm{fact}} &=& 7.5 \mbox{~MeV} , \hspace{3em}
|f^\varphi (a_1^{\mathrm{fact}}+a_1^{\mathrm{corr}}) \,|_{\pi(1300)}
  \hspace{0.55em} \,=\; 
              7.4 \mbox{~or~} 7.6 \mbox{~MeV} , \nonumber \\
f_{K_0^*(1430)}\, a_1^{\mathrm{fact}} &=& \; 48 \mbox{~MeV} , \hspace{3em}
|f^\varphi (a_1^{\mathrm{fact}}+a_1^{\mathrm{corr}}) \,|_{K_0^*(1430)}
  \,=\;       \; 47 \mbox{~or~} \, 48 \mbox{~MeV} .
\label{coeffs-D}
\end{eqnarray}
Notice the enormous correction to the naive factorization result for
the $a_0(1450)$. The impact of nonfactorizing corrections is similarly
strong for the $a_0(980)$ if we take the minimum value
$f_{a_0(980)}=0.7$~MeV from Eq.~(\ref{nar}). For the $\pi(1300)$,
nonfactorizing terms remain always moderate since $|f^\varphi
a_1^{\mathrm{corr}}|_{\pi(1300)}$ is small even compared with the
lowest estimate of $f_{\pi(1300)}$ in Eq.~(\ref{heavy-pi}). The
$K_0^*$ has a decay constant much larger than $|f^\varphi
a_1^{\mathrm{corr}}|_{K_0^*}$ and is the only case where the
corrections hardly matter.

Using ${\cal{B}}(B^0 \to D^- \pi^+)=(3.0 \pm 0.4)\cdot 10^{-3}$ from
\cite{Groom:2000in} and the maximal values in (\ref{coeffs-D}), we
obtain the branching ratios in Table~\ref{tab:branch}. To show their
dependence on the choice of renormalization scale, we give them for
$\mu=m_b$ and $\mu=m_b/2$. We take the latter as an indication of how
large the branching ratios can be in QCD factorization, although they
are not upper bounds in a rigorous sense. We also show the
corresponding results from naive factorization, where for consistency
of comparison we have again neglected the corrections discussed below
Eq.~(\ref{eq:rate}). Here the scale dependence is minute, less than 2
percent, and the values in Table~\ref{tab:branch} are those for
$\mu=m_b$.

\begin{table}[tb]
\begin{center}
\renewcommand{\arraystretch}{1.2}
\begin{tabular}{lccc} \hline \hline
decay mode & naive factorization & \multicolumn{2}{c}{QCD factorization}
\\ 
 & & ~~~$\mu=m_b$~~~ & ~~~$\mu=m_b/2$~~~ \\ \hline
$\bar{B}^0 \to D^+ a_0(980)$ & $1.1\cdot 10^{-6}$
                             & $2.0\cdot 10^{-6}$ & $4.0\cdot 10^{-6}$
\\
$\bar{B}^0 \to D^+ a_0(1450)$ & $8.6\cdot 10^{-8}$
                              & $5.8\cdot 10^{-7}$ & $2.1\cdot 10^{-6}$
\\
$\bar{B}^0 \to D^+ b_1,\, D^+ a_2$ & 0
                              & $3.5 \cdot 10^{-7}$ & $1.7 \cdot 10^{-6}$
\\
$\bar{B}^0 \to D^+ \pi(1300)$ & $9.1\cdot 10^{-6}$
                              & $9.3\cdot 10^{-6}$ & $9.6\cdot 10^{-6}$
\\
$\bar{B}^0 \to D^+\pi_2,\, D^+\rho_3$ & 0
                              & $1.4 \cdot 10^{-9}$ & $8.1 \cdot 10^{-9}$
\\
$\bar{B}^0 \to D^+ K^*_0(1430)$ & $2.0\cdot 10^{-5}$ & $2.0\cdot 10^{-5}$
                                & $2.1\cdot 10^{-5}$
\\
$\bar{B}^0 \to D^+ K^*_2$ & 0 & $1.9\cdot 10^{-8}$ & $9.2\cdot 10^{-8}$
\\ \hline
$\bar{B}^0 \to D^{*+} a_0(980)$ & $1.0\cdot 10^{-6}$
                                & $1.8\cdot 10^{-6}$ & $3.7\cdot 10^{-6}$
\\
$\bar{B}^0 \to D^{*+} a_0(1450)$ & $7.9\cdot 10^{-8}$
                                 & $5.2\cdot 10^{-7}$ & $1.9\cdot 10^{-6}$
\\
$\bar{B}^0 \to D^{*+} b_1,\, D^{*+} a_2$ & 0
                                 & $2.9 \cdot 10^{-7}$ & $1.5 \cdot 10^{-6}$
\\
$\bar{B}^0 \to D^{*+} \pi(1300)$ & $8.3\cdot 10^{-6}$
                                 & $8.4\cdot 10^{-6}$ & $8.4\cdot 10^{-6}$
\\
$\bar{B}^0 \to D^{*+}\pi_2,\, D^{*+}\rho_3$ & 0
                                 & \, $5.7\cdot 10^{-10}$ & $3.2\cdot 10^{-9}$
\\
$\bar{B}^0 \to D^{*+} K^*_0(1430)$ & $1.8\cdot 10^{-5}$
                                   & $1.9\cdot 10^{-5}$ & $1.9\cdot 10^{-5}$
\\
$\bar{B}^0 \to D^{*+} K^*_2$ & 0 & $1.5\cdot 10^{-8}$ & $7.7\cdot 10^{-8}$
\\ \hline \hline
\end{tabular}
\end{center}
\caption{\label{tab:branch}Branching ratios for various decay modes
obtained in QCD factorization for two choices of the renormalization
scale $\mu$. For comparison we also give the branching ratios in the
naive factorization approach.  We recall that we have set $f_{b_1}=0$
for lack of better knowledge.}
\end{table}

We proceed to decays into a vector meson, $\bar{B}^0 \to D^{*+} X^-$.
The contraction of $q_\mu$ with the matrix element $\langle D^{*}
|\bar{c}\gamma^\mu\gamma_5 b |\bar{B}\rangle $ depends again only on a
single form factor, commonly referred to as $A_0$ and defined e.g.\ in
\cite{Ali:1998eb}. The decay rate is then given by
\begin{eqnarray}
\Gamma(\bar{B}^0 \to D^{*+} X^-) &=& \frac{G_F^2}{4 \pi}\,
p_X^3\, |V^{\phantom{*}}_{cb} V_{ud}^*|^2\, 
\Big|f^\varphi (a_1^{\mathrm{fact}}+
   a_1^{\mathrm{corr}})\,  A_0(m_X^2) \,\Big|^2 .
\end{eqnarray}
As mentioned in Sect.~\ref{sec:heavy-light-qcd}, the coefficients
$a_1^{\mathrm{corr}}$ for decays into $D^*$ are different from those
into $D$, and we now have
\begin{eqnarray}
|f^\varphi (a_1^{\mathrm{fact}}+a_1^{\mathrm{corr}}) \,|_{a0(980)}
  &=& 2.2 \mbox{~or~} 3.5 \mbox{~MeV} , \nonumber \\
|f^\varphi (a_1^{\mathrm{fact}}+a_1^{\mathrm{corr}}) \,|_{a0(1450)}
  &=& 1.2 \mbox{~or~} 1.9 \mbox{~MeV} , \nonumber\\
|f^\varphi (a_1^{\mathrm{fact}}+a_1^{\mathrm{corr}}) \,|_{\pi(1300)}
  &=& 7.5 \mbox{~MeV},
\nonumber \\
|f^\varphi (a_1^{\mathrm{fact}}+a_1^{\mathrm{corr}}) \,|_{K_0^*(1430)}
  &=& 47 \mbox{~or~} 49 \mbox{~MeV} 
\label{coeffs-D-star}
\end{eqnarray}
at $\mu=m_b$. The analogous expression for the ratio (\ref{eq:rate})
of decay rates still holds, although with different mass corrections.
We estimate them to be not much larger than for the $D$ and neglect
them as before. Normalizing to ${\cal{B}}(B^0 \to D^{*-} \pi^+)=(2.76
\pm 0.21)\cdot 10^{-3}$ from \cite{Groom:2000in} we obtain the
branching ratios in Table~\ref{tab:branch}, which are somewhat smaller
than the corresponding ones for decays into the $D$. With the
exception of decays into $K^*_2, \pi_2$, or $\rho_3$, all estimated
branching ratios are larger than $10^{-7}$ and within experimental
reach at the $B$-factories. We expect similar branching ratios for
$B_s$ decays into $D_s^{(*)}$ and the same candidate mesons $X$, up to
SU(3) breaking effects.

To facilitate comparison of the branching ratios of
Table~\ref{tab:branch} into $K_0^*$ and $K_2^*$ with those into a $K$,
we estimate the latter as
\begin{eqnarray}
{\cal{B}}(\bar{B}^0 \to D^+ K^-) &\simeq& \frac{f^2_K}{f_\pi^2}
\left| \frac{V_{us}}{V_{ud}} \right|^2
{\cal{B}}(\bar{B}^0 \to D^+ \pi^-)=
(2.4 \pm 0.3) \cdot 10^{-4} \, ,
 \nonumber \\
{\cal{B}}(\bar{B}^0 \to D^{*+} K^-) & \simeq& \frac{f^2_K}{f_\pi^2}
\left| \frac{V_{us}}{V_{ud}} \right|^2
{\cal{B}}(\bar{B}^0 \to D^{*+} \pi^-)=(2.2 \pm 0.2) \cdot 10^{-4} \, ,
\end{eqnarray}
which is in good agreement with data on the ratios
${\cal{B}}(\bar{B}^0 \to D^{+} K^-)/{\cal{B}}(\bar{B}^0 \to D^{+}
\pi^-)=0.079 \pm 0.011$
and
${\cal{B}}(\bar{B}^0 \to D^{*+} K^-)/{\cal{B}}(\bar{B}^0 \to D^{*+}
\pi^-)= 0.074 \pm 0.016$
reported in \cite{Iijima:2001gz}.

A considerable source of uncertainty in our decay rate estimates are
the unknown meson decay constants and distribution amplitudes. Let us
illustrate this for the channel $\bar{B}^0 \to D^+ a_0(980)$. With the
minimal value of $f_{a_0(980)}$ in Eq.~(\ref{malt}) we obtain
$|f^\varphi (a_1^{\mathrm{tree}}+a_1^{\mathrm{corr}})|_{a0(980)} =
\mbox{1.3 or 1.9~MeV}$ at $\mu=m_b$. The larger of the two
possibilities corresponds to a branching ratio ${\cal{B}}(\bar{B}^0
\to D^+ a_0(980))= 5.8 \cdot 10^{-7}$, about three times smaller than
the corresponding one in Table~\ref{tab:branch}.

We have already seen the sensitivity of our results on the choice of
renormalization scale. This uncertainty is largest for the branching
ratios which are zero in naive factorization, with a variation by a
factor of 5 to 6 between $\mu=m_b$ and $\mu=m_b/2$. How important it
is for the other channels depends on the actual size of the decay
constants and distribution amplitudes. We also remark that the
reliability of the QCD factorization approach for our ``light'' mesons
with masses in the range from 1 to 1.7~GeV might be questionable, at
least unless finite-mass corrections can be taken into account.

Most important, however, is that the hard nonfactorizing contributions
are small on the scale of the amplitudes for unsuppressed decays. With
all uncertainties discussed above we find that $f^\varphi
a_1^{\mathrm{corr}}$ is at most a few MeV, i.e., less than 5\% of
$f_\pi a_1^{\mathrm{fact}}$. One may well expect that soft corrections
(or annihilation graphs when they can occur) are bigger than the hard
ones. For all our meson candidates except the $K_0^*$ they would then
lead to considerably larger branching ratios than we have
estimated. 

The decays into $K_0^*$ are different in this context. Here the
perturbative corrections to the prediction of naive factorization are
quite small and their uncertainties less relevant, and further soft
corrections might or might not overshadow the factorizing piece. Once
the decay constant of the $K_0^*$ is known experimentally, one should
of course refine our estimates by taking into account the phase space
and form factor corrections in (\ref{eq:rate}).

%------------------------------------------------------------------
\subsection{Background from $B^0$ decay}

\label{sec:background}

We see in Fig.~\ref{fig:tree} that the same final state of our signal
mode $\bar{B}^0 \to D^+ X^-$ just discussed can be produced in the
decay of the $C\!P$ conjugated parent meson, $B^0 \to D^+ X^-$. As
mentioned in Sect.~\ref{sec:flavor}, this background is CKM suppressed
with respect to the signal.  On the other hand, the signal mode is
punished by a small decay constant, whereas the background goes with
$f_D\sim 200$~MeV.  One expects the background-to-signal ratio to be
large, since at the amplitude level $\lambda^2 f_D/f_X \sim
{\cal{O}}(1)$. We recall that no such background exists in decays into
a strange final state, $\bar{B}^0 \to D^+ X_s^-$ or $\bar{B}_s \to
D_s^+ X^-$.

The background to $\bar{B}^0 \to D^+ X^-$ can of course be removed by
flavor tagging. Experimental discrimination between $B$ and $\bar{B}$
is however challenging in decays with branching ratios of less than
$10^{-5}$, and it is worthwhile to see how far one can go without a
flavor tag.
Let us therefore investigate in more detail the branching ratios of
the background decays for the case of the $a_0$.  We parameterize the
matrix element for the $B \to a_0$ transition in terms of form factors
$F_0^a$ and $F_1^a$ as
\begin{eqnarray}\label{eq:ffba0}
\langle a_0(p^\prime) | \bar u \gamma^\mu  \gamma_5 b | 
                                               \bar{B}(p)\rangle  
&=&  F^a_1(q^2) \left\{
(p+p^\prime)^\mu - \frac{m_B^2-m_{a_0}^2}{q^2}\, q^\mu \right\} +
F^a_0(q^2)\, \frac{m_B^2-m_{a_0}^2}{q^2}\, q^\mu . \hspace{1em}
\end{eqnarray}
Assuming naive factorization, the decay rates for $B^0 \to D^{(*)+}
a_0^-$ decays can be written as
\begin{eqnarray}
\Gamma(B^0 \to D^+ a_0^-)&=&
\frac{G_F^2}{16 \pi}\, \frac{(m_B^2-m_{a_0}^2)^2}{m_B^2}\, p_{a_0}\,
\Big| V^{\phantom{*}}_{cd} V_{ub}^* \Big|^2 \,
(a_1 f_D)^2\, |F^a_0(m_D^2)|^2 ,
\nonumber \\
\Gamma(B^0 \to D^{*+} a_0^-)&=&
\frac{G_F^2}{4 \pi}\, p_{a_0}^3\,
\Big| V^{\phantom{*}}_{cd} V_{ub}^* \Big|^2 \,
(a_1 f_{D^*})^2\, |F^a_1(m_{D^*}^2)|^2 ,
\end{eqnarray}
where $a_1$ is the universal coefficient for color allowed decays in
naive factorization, introduced in Sect.~\ref{sec:fac}. Using
$a_1=1.03$ we find
\begin{eqnarray}
{\cal{B}}(B^0 \to D^+ a_0(980))&=&2.1 \cdot 10^{-6}  
\left( \frac{|V^{\phantom{*}}_{cd} V_{ub}^*|}{
             7.3 \cdot 10^{-4}} \right)^2
\left( \frac{f_D}{200 \mbox{~MeV}} \right)^2 
\left( \frac{F^a_0(m_D^2)}{0.5} \right)^2 
\frac{ \tau_{B^0}}{1.55 \mbox{~ps}} , \nonumber \\
{\cal{B}}(B^0 \to D^{*+} a_0(980))&=&1.9 \cdot 10^{-6} 
\left( \frac{|V^{\phantom{*}}_{cd} V_{ub}^*|}{
             7.3 \cdot 10^{-4}} \right)^2
\left( \frac{f_{D^*}}{230 \mbox{~MeV}} \right)^2 
\left( \frac{F^a_1(m_{D^*}^2)}{0.5} \right)^2 
\frac{ \tau_{B^0}}{1.55 \mbox{~ps}} , \hspace{2em}
\label{eq:bgd}
\end{eqnarray}
where we have indicated the sensitivity to several input parameters
which at present have significant un\-certainties.  In particular,
very little is known about the form factors for $B \to X$
transitions. Chernyak \cite{Chernyak:2001hs} has recently estimated
the form factor $F_1(0)^{B \to a_0(1450)} \simeq 0.46$ with light-cone
sum rules, indicating only a small enhancement over the corresponding
one into a pion, where he cites $F_1(0)^{B \to \pi} \simeq 0.3$. That
the form factor for $B\to a_0$ should rather be larger than the one
for $B\to \pi$ is also plausible in the Bauer-Stech-Wirbel approach
\cite{Wirbel:1985ji}. To see this, consider the constituent $q\bar{q}$
wave function of a charged $a_0$. In the case where the $q$ and
$\bar{q}$ spins couple to $S_3=0$, it has a zero at momentum fraction
$u=\frac{1}{2}$ due to charge conjugation, up to small isospin
breaking effects. Being normalized to one, this wave function is then
more pronounced towards the endpoints $u=0$ and $u=1$, and thus can
have a greater overlap with the asymmetric wave function of the $B$
than the pion wave function can. Further information may be obtained
in relativistic quark models \cite{Isgur:1989gb}.

We wish to point out that experimental information on the form factor
$F^a_1(m_D^2)$ can be obtained from semileptonic decays at
$q^2=m_D^2$,
\begin{eqnarray}
\frac{d\Gamma(\bar{B}^0 \to a_0^+\, \ell^- \bar{\nu}_\ell)}{dq^2}&=&
\frac{G_F^2}{24 \pi^3} \,
p_{a_0}^3 \, |V_{ub}|^2\, |F^a_1(q^2)|^2 ,
\end{eqnarray}
where we expect similar statistics as for the semileptonic decays $B
\to \pi,\rho$ with branching ratios of a few $10^{-4}$, if the form
factors have comparable size.
One may then relate $F^a_0$ with $F^a_1$ using large energy effective
theory (LEET) \cite{Charles:1999dr}, originally introduced in
Ref.~\cite{Dugan:1991de}. We are in the kinematical situation where a
light meson (here the $a_0$) is emitted from a heavy parent with large
recoil $q^2=m_{D}^2\ll m_b^2$ and an energy $E=(m_B^2-q^2+m_{a0}^2)/(2
m_B)$ much larger than $\Lambda_{QCD}$ and the light masses in the
process.  This is the region of applicability of LEET. To leading
order in $1/E$ and $1/m_b$, we derive
\begin{eqnarray}
F^a_1(E,m_b)=\frac{m_B}{2 E}\, F^a_0(E,m_b) .
\end{eqnarray}
This is the analog of Eqs.~(104) and (105) in \cite{Charles:1999dr},
to which we refer for details. 
Hence, the form factors are equal to leading order in the
large energy limit. 

{}From the branching ratios in Eq.~(\ref{eq:bgd}) we conclude that
rather likely the background from decays of $B^0$ mesons into $a_0$
does not overshadow the signal. A similar discussion can be given for
decays into the other $I=1$ mesons of Table~\ref{tab:friends}.
Assuming naive factorization and no anomalous behavior of the relevant
form factors, we quite generally expect branching ratios of the
background modes $B^0 \to D^{(*)+} X^-$ of order $10^{-6}$. Any
significant excess over both this and the branching ratios given in
Table~\ref{tab:branch} would imply that either there are important
nonfactorizing contributions in the signal, or that naive
factorization drastically fails in the background channel.

%%%%%%%%%%%%%%%%%%%%%%%%%%%%%%%%%%%%%%%%%%%%%%%%%%%%%%%%%%%%%%%%%%%
\section{Summary}
\label{sec:sum}
%%%%%%%%%%%%%%%%%%%%%%%%%%%%%%%%%%%%%%%%%%%%%%%%%%%%%%%%%%%%%%%%%%%

We have explored how to obtain quantitative information on
nonfactorizing effects in exclusive $b$ decays, using channels where
such contributions are not hidden behind larger factorizing pieces. We
achieve this through ``switching off'' the factorizing contribution by
choosing final-state mesons with either a small decay constant or spin
$J\ge 2$.  Our proposal is similar in spirit to the study of decay
channels where the quark content of the final state does not admit
factorizing contributions, such as $B^0 \to K^+ K^-$
\cite{Gronau:1998gr}, $B_s \to \pi^+ \pi^-$, $\pi^0 \pi^0$, or
$b$-decays into baryon-antibaryon pairs (see
e.g.~\cite{Neubert:1997uc} and references therein). Suppression of the
factorizable contributions thus highlights factorization breaking
effects, such as annihilation graphs, soft or hard interactions, and
in general any mechanism dominated by long-distance physics, which
disconnects the $b$ decay vertex from the final state meson. We have
explicitly shown that hard nonfactorizing contributions, calculated in
the QCD factorization framework, can yield sizeable contributions to
the decay amplitude.

In a systematic study, compiled in Tables~\ref{tab:flavor},
\ref{tab:flavor-two}, and \ref{tab:baryons}, we have shown that our
method applies to a variety of mesons and channels, in decays of
$B_{u,d}$, $B_s$, and $b$ baryons.  In particular, the mesons $X$ we
have selected cover a wide range of masses, which makes it possible to
explore whether the energy-mass ratio $E_X /m_X$ in the parent rest
frame is a relevant parameter to ensure factorization, as is suggested
by color transparency but not by large $N_c$ arguments
\cite{Dugan:1991de}.

We have presented a detailed analysis of color allowed decays $B\to
D^{(*)} X$ and $B_s^{\phantom{*}}\to D_s^{(*)} X$ with a light meson
$X$. When the factorizable contribution is suppressed, e.g.\ for the
scalar $a_0$, we found that hard nonfactorizing corrections can be of
similar magnitude or even larger than the Born term. They remain
however much smaller than the amplitudes of corresponding
nonsuppressed decays, for instance into a $\pi$.  In several cases we
found branching ratios substantially enhanced over the ones calculated
in the naive factorization approach, see Table~\ref{tab:branch}, and
are within the reach of existing and future experiments at the
$B$-factories BaBar, Belle, CLEO and at hadron colliders like the
Tevatron and the LHC. Comparison of these decays with modes where the
ejected meson is a $D$ meson and further study of those into charmonia
$\chi_{c0}$ or $\chi_{c2}$ should give complementary information on
the origin and limitations of the factorization approach.

$B$ decays with light-light final states are more complex. Modes such
as $\bar{B}^0\to a_0^+ a_0^-$ and $B_s\to K^+ a_0^-$ are not entirely
suppressed by the small decay constant of the $a_0$ due to the
presence of scalar penguin operators, but we find the corresponding
amplitudes to be much smaller than those of $\bar{B}^0\to \pi^+ \pi^-$
or $B_s\to K^+ \pi^-$. Such factorizing penguin contributions can be
eliminated altogether with higher-spin mesons like the $b_1$ or $a_2$
instead of the $a_0$. We expect hard nonfactorizing contributions to
be moderate, too, so that experimental information on such decays
could again tell us whether nonperturbative effects are large.
For penguin dominated decays like $\bar{B}^0\to \pi^+ K_2^{*-}$ the
situation is less clear-cut on the quantitative level, but we argue
that they can give valuable indications on the importance of penguin
annihilation contributions. This is a particularly controversial issue
since different conclusions regarding such decays have been drawn in
the QCD factorization and the PQCD scenarios
\cite{Beneke:2001ev,Keum:2001wi}.

To conclude, we find that the $b$ decays presented here provide a tool
for studying important issues in exclusive nonleptonic decays. We
stress that in order to make this tool more quantitative, the decay
constants of the $a_0(980)$, $a_0(1450)$, $\pi(1300)$, $K_0^*(1430)$
and $b_1$ mesons should be known experimentally. Their determination
from $\tau$ decays should be in reach of the existing experiments at
BaBar, Belle and CLEO, and even more of dedicated $\tau$-charm
factories. Information on the distribution amplitudes of $a_0(980)$,
$a_0(1450)$, $a_2$, $\pi(1300)$, $\pi_2$, which are needed for the
calculation of hard nonfactorizable contributions, could be obtained
from $\gamma^*\gamma$ collisions at the $B$ factories.

\vspace{0.4cm}

{\bf  {Note added}}: 
The suppression mechanisms discussed in this work provide
opportunties to study $C\!P$ violation in various $B$ decays into
designer mesons $a_0$, $b_1$, $a_2$, etc.  This has been explored
independently in two recent studies \cite{slac} and \cite{berkeley}.

%%%%%%%%%%%%%%%%%%%%%%%%%%%%%%%%%%%%%%%%%%%%%%%%%%%%%%%%%%%%%%%%%%%
\section*{Acknowledgments}
%%%%%%%%%%%%%%%%%%%%%%%%%%%%%%%%%%%%%%%%%%%%%%%%%%%%%%%%%%%%%%%%%%%

It is a pleasure to thank P.~Ball, S.~J.~Brodsky, G.~Buchalla,
H.~G.~Dosch, A.~Kagan, H.-n.~Li, M.~Neubert, S.~Spanier, and H.~Quinn
for discussions, and V.~Braun for correspondence. We also thank A.~Ali
for his careful reading of the manuscript.

This work was initiated when M.~D.\ was visiting SLAC. He acknowledges
financial support through the Feodor Lynen Program of the Alexander
von Humboldt Foundation, and thanks the SLAC theory group for its
hospitality.

%%%%%%%%%%%%%%%%%%%%%%%%%%%%%%%%%%%%%%%%%%%%%%%%%%%%%%%%%%%%%%%%%%%
\section*{Appendix}
%%%%%%%%%%%%%%%%%%%%%%%%%%%%%%%%%%%%%%%%%%%%%%%%%%%%%%%%%%%%%%%%%%%

In this appendix we estimate the size of the leading-twist
distribution amplitude $\varphi$ of several mesons. Our method is
based on the connection between distribution amplitudes and the Fock
state expansion in QCD, and closely follows the discussion in
\cite{Lepage:1983}, to which we refer for details. The starting point
is to decompose a hadron state on Fock states consisting of current
quarks and gluons, $q\bar{q}$, $q\bar{q}g$, etc. The coefficients in
this expansion are the light-cone wave functions for each parton
configuration. For a $d\bar{u}$ meson, one has
\begin{equation}
|X^-\rangle_{J_3=0} = \int \frac{du}{\sqrt{u(1-u)}}\,
   \frac{d^2k_\perp}{16\pi^3}\,
   \frac{ |d_\uparrow\, \bar{u}_\downarrow\rangle \pm
          |d_\downarrow\, \bar{u}_\uparrow\rangle }{\sqrt{2}}\,
   \psi(u,k_\perp)\, 
  + \ldots ,
\end{equation}
where $u$ and $k_\perp$ denote the light-cone momentum fraction and
transverse momentum of the $d$ quark in the meson. The arrows indicate
quark and antiquark helicities, and the $+$ and $-$ respectively apply
to mesons with natural and unnatural parity. The states $|d_\uparrow\,
\bar{u}_\downarrow\rangle$ and $|d_\downarrow\,
\bar{u}_\uparrow\rangle$ are understood to be coupled to color
singlets. By $\ldots$ we have denoted the Fock states $|d_\uparrow\,
\bar{u}_\uparrow\rangle$ and $|d_\downarrow\,
\bar{u}_\downarrow\rangle$ with aligned quark helicities, and Fock
states with additional partons. The connection of the light-cone wave
function $\psi(u,k_\perp)$ with the distribution amplitude defined in
Eq.~(\ref{da-def}) is
\begin{equation}
\int \frac{d^2k_\perp}{16\pi^3}\, \psi(u,k_\perp) = 
\frac{1}{2\sqrt{6}}\, \varphi(u) .
\label{lcwf-da}
\end{equation}
The probability to find the $d\bar{u}$ Fock state with antialigned
helicities in the meson $X$ is
\begin{equation}
P = \int du\, \frac{d^2k_\perp}{16\pi^3}\, | \psi(u,k_\perp) |^2 .
\label{proba}
\end{equation}
This should be below 1 since it is the probability to find a current
$q\bar{q}$ pair in the meson, without further gluons or sea quark
pairs. Note that this is different from the $q\bar{q}$ wave functions
in constituent quark models, which are by definition normalized to
1. Let us now use the relation (\ref{proba}) to estimate the size of
$\varphi(u)$. In order to achieve this, we need to make an ansatz for
the $k_\perp$ dependence. A form consistent with several theoretical
requirements \cite{Lepage:1983,Chibisov:1995ss} is
\begin{equation}
\psi(u,k_\perp) = \frac{16\pi^2 a^2}{u(1-u)} 
  \exp\left[-\frac{a^2 k_\perp^2}{u(1-u)}\right]\, 
\frac{1}{2\sqrt{6}}\,\varphi(u) ,
\label{exp-ansatz}
\end{equation}
where the prefactor of the exponential is imposed by the
normalization (\ref{lcwf-da}). $a$ plays the role of a transverse
size parameter of the $d\bar{u}$-pair in the meson. For the pion,
Brodsky and Lepage \cite{Lepage:1983} obtained $a_\pi\approx
0.86$~GeV$^{-1}$ with the above ansatz and the asymptotic form
$\varphi_\pi(u)=f_\pi\, 6u(1-u)$ for the distribution amplitude. This
corresponds to an average transverse momentum $\langle k_\perp^2
\rangle \approx (370$~MeV)$^2$ and to a Fock state probability of
$P_\pi \approx 0.25$. We will take the same values for the mesons we
discuss here, which is certainly a crude assumption but should give
the correct order of magnitude.

With the ansatz (\ref{exp-ansatz}) and the Gegenbauer expansion
(\ref{gegenbau}) we obtain
\begin{equation}
P = 2 \pi^2 (a f^\varphi)^2 \left( B_0^2 + \sum_{n=1}^\infty
         \frac{3(n+2)(n+1)}{2(2n+3)}\, B_n^2\, \right) .
\label{proba-gegen}
\end{equation}
Consider now a meson for which $B_0=0$, such as the $a_2$ or
$K_2^{*}$. If we take $a=a_\pi$ and $P=P_\pi \approx 0.25$ and retain
only the term with $B_1$ in the Gegenbauer expansion, we obtain
\begin{equation}
|f^\varphi B_1| \approx 100~\mbox{MeV} .
\label{b1-est}
\end{equation}
Including the zeroth term $f^\varphi B_0$ in the Gegenbauer expansion,
as is appropriate for the charged $a_0$, $K_0^*$ and $b_1$, would
decrease this estimate by about $5\%$ for the $K_0^*$ when taking
$f_{K_0^*}=42$~MeV. The effect of that term for the $a_0$ or $b_1$ can
be neglected even more safely.

In order to explore the dependence of our estimate on the ansatz we
made for the $k_\perp$ dependence of the wave function, we take an
alternative form
\begin{equation}
\psi(u,k_\perp) = \frac{16\pi^2 \tilde{a}^4}{u^2(1-u)^2}\, k_\perp^2
   \exp\left[-\frac{\tilde{a}^2 k_\perp^2}{u(1-u)}\right]\, 
\frac{1}{2\sqrt{6}}\,\varphi(u) ,
\label{alt-ansatz}
\end{equation}
which has a node at $k_\perp=0$. For the Fock state probability we
find the same expression as (\ref{proba-gegen}) with $a$ replaced by
$\tilde{a} /\sqrt{2}$. Choosing $\tilde{a}=a$ we then get an estimate
of $|f^\varphi B_1|$ larger by a factor $\sqrt{2}$. If instead one
requires the average $\langle k_\perp^2 \rangle$ to be the same with
the two forms (\ref{exp-ansatz}) and (\ref{alt-ansatz}), one finds
$\tilde{a} = \sqrt{3} a$ and thus an estimate of $|f^\varphi B_1|$
smaller by a factor of $\sqrt{2/3}$. Given these observations we
expect that (\ref{b1-est}) should give the correct order of magnitude
of the first Gegenbauer coefficient.

We should add that this does not hold for mesons that are not
$q\bar{q}$ bound states in the constituent quark picture but for
instance made from $q\bar{q}q\bar{q}$, which may be the case for one
of the $a_0$ mesons. Is plausible that for such a system the
probability of finding a single current $q\bar{q}$ pair in this meson
is reduced compared with the one of a conventional $q\bar{q}$
state. Correspondingly, its twist-two distribution amplitude $\varphi$
and the coefficients $f^\varphi B_n$ would be smaller than estimated
here.

Our considerations are easily adapted to the case of mesons where $B_1$
is zero or isospin suppressed, such as the $\pi(1300)$, the $\pi_2$,
or the $\rho_3$. Retaining only $B_2$ in the Gegenbauer expansion of
$\varphi(u)$ and taking as before $a \approx a_\pi$ and $P=P_\pi
\approx 0.25$, we obtain
\begin{equation}
|f^\varphi B_2| \approx 80~\mbox{MeV} .
\label{b2-est}
\end{equation}

So far we have not displayed the dependence of both the distribution
amplitude $\varphi$ and the light-cone wave function $\psi$ on the
factorization scale $\mu$, which physically represents the resolution
scale of the $q\bar{q}$ pair. Our above estimates are understood as
corresponding to a hadronic scale, say, $\mu=1$~GeV. Evolving up to
$\mu=m_b$ we obtain $|f^\varphi B_1| \approx 75$~MeV from
Eq.~(\ref{b1-est}) and $|f^\varphi B_2| \approx 50$~MeV from
Eq.~(\ref{b2-est}). 

To conclude this section, we wish to point out that experimental
constraints on the distribution amplitudes for the neutral mesons
$X=a_0$, $a_2$, $\pi(1300)$, $\pi_2$ can be obtained from the process
$\gamma^*\gamma\to X$ at virtualities $Q^2$ of the photon much larger
than the meson mass. This can be measured in $e^+e^-\to e^+e^- X$, and
the CLEO data for $X=\pi,\eta,\eta'$ are in fact one of our best
sources of information on the corresponding distribution amplitudes
\cite{Gronberg:1998fj}. To leading order in $1/Q^2$ and in $\alpha_s$,
the amplitude for $\gamma^*\gamma\to X$ is proportional to $f^\varphi
(B_0 + B_2 + B_4 + \ldots)$ for mesons with unnatural parity, and to
$f^\varphi (B_1 + B_3 + B_5 + \ldots)$ for mesons with natural
parity. According to our above estimates, one then expects cross
sections comparable to the one for $\pi$ production, so that the
measurement of these reactions at large $Q^2$ may well be in the reach
of the $B$ factories.

%%%%%%%%%%%%%%%%%%%%%%%%%%%%%%%%%%%%%%%%%%%%%%%%%%%%%%%%%%%%%%%%%%%

\end{document}